\documentclass[fleqn,10pt]{wlscirep}
\usepackage[utf8]{inputenc}
\usepackage[T1]{fontenc}

\usepackage{lipsum}
\usepackage{bbold}
\usepackage{bbm}
\usepackage{graphicx}
\usepackage{latexsym,amsmath,verbatim}
\usepackage{color}
\usepackage{rotating}
\usepackage{multirow}
\usepackage[english]{babel}
\usepackage{todonotes}
\usepackage{etoc}

\definecolor{snorkelBlue}{rgb}{0,0.31,0.52}
\definecolor{peachColor}{rgb}{0.97,0.47,0.42}
\definecolor{Biscaybay}{rgb}{0, 0.556, 0.655}
\definecolor{lightGray}{rgb}{0.7,0.7,0.7}

\newcommand{\humI}{\rho^{H}}
\newcommand{\mosI}{\rho^{M}}
\newcommand{\lm}{\lambda^{HM}}
\newcommand{\lh}{\lambda^{MH}}

\title{Vector-borne epidemics driven by human mobility}
\author[1,2]{David Soriano-Pa\~nos}
\author[3]{Judy Heliana Arias-Castro} 
\author[3]{Hector J. Mart\'{\i}nez}
\author[4]{Sandro Meloni}
\author[1,2,*]{Jes\'us G\'omez-Garde\~nes}

\affil[1]{GOTHAM Lab., Institute for Biocomputation and Physics of Complex Systems (BIFI), University of Zaragoza, 50018 Zaragoza, Spain}
\affil[2]{Departament of Condensed Matter Physics, University of Zaragoza, 50009 Zaragoza, Spain}
\affil[3]{Department of Mathematics, Universidad del Valle, 760032 Santiago de Cali, Colombia}
\affil[4]{Instituto de F\'{\i}sica Interdisciplinar y Sistemas Complejos IFISC (CSIC-UIB), Campus UIB, 07122 Palma de Mallorca, Spain}
\affil[*]{gardenes@unizar.es}

\begin{abstract}
Vector-borne epidemics are the result of the combination of different factors such as the crossed contagions between humans and vectors, their demographic distribution and human mobility among others.  The current availability of information about the former ingredients demands their incorporation to current mathematical models for vector-borne disease transmission. Here, relying on metapopulation dynamics, we propose a framework whose results are in fair agreement  with those obtained from mechanistic simulations. This framework allows us to derive an expression of the epidemic threshold capturing with high accuracy the conditions leading to the onset of epidemics. Driven by these insights, we obtain a prevalence indicator to rank the patches according to the risk of being affected by a vector-borne disease. We illustrate the utility of this epidemic risk indicator by reproducing the spatial distribution Dengue cases reported in the city of Santiago de Cali (Colombia) from 2015 to 2016.

\end{abstract}
\begin{document}

\maketitle

\section*{Introduction} 

The explosive dissemination of Zika virus across the Americas has been one of the major concerns of public health organizations across the world in the recent years \cite{WHO16}. Zika's global threat is, unfortunately, the last example of the extremely rapid dissemination of mosquito-borne flaviviruses over the past two decades. From Dengue to Zika,  through West Nile and Chikungunya viruses, more than one billion people are infected and more than one million people die from vector-borne diseases (VBD) every year \cite{WHO14}. According to the World Health Organization (WHO), VBD are responsible of one sixth of the illness worldwide and more than half of human population live in risk areas for these diseases \cite{WHO2004}. 
Moreover, the threat of new emergent VBD in tropical and ecuatorial regions progressively span across more temperate areas as a byproduct of climate change. As temperature rises, the areas that are conducive to mosquitoes expand, meaning more opportunities for VBD to spread \cite{Kraemer}. For the case of Malaria, an increase of global temperature of 2-3 degrees would rise the population at risk by several hundred million people while, in the case of Dengue, the population at risk would pass from 1-2 billion people in 1990 to 5-6 billion in 2085 \cite{Reiter,Hales,Reiter08}. 
  
The lack of efficient vaccines constitutes another significant bump on the road of facing flaviviruses epidemics. Despite the efforts for finding effective immunization means to slow down the advance of {\em Aedes}-borne viruses, the most common way for preventing outbreaks is the use of pesticides, larvivorous fishes or Wolbachia bacteria, all of them directly acting over the vector \cite{Control}.
The use of geolocalized control strategies, however, seems not to be an effective prevention strategy when facing the threat of a global-scale pandemic. On the contrary, the fast transcontinental movement of VBD demands a coordinated action of the involved actors for the efficient use of local control means. This implies taking into account that those populations at risk are not isolated and, as for human-human transmission diseases \cite{Colizza06}, human mobility plays a key role in the spread of VBD across different populations \cite{Adams}. In this sense, over the last decades, the increasing use and efficiency of transportation means have led to an explosion of human mobility, including urban, regional and long-range displacements \cite{airline,marta1,gourab1,gourab2,ernesto1}. Despite the social and economical benefits, the expansion of human mobility networks produced, as a byproduct, the speed up of epidemic waves and the emergence of correlated  outbreaks  in far away regions. For this reason,  incorporating human mobility into disease transmission models has become a must when proposing mathematical frameworks aimed at capturing the contagions patterns observed in actual epidemic scenarios. 

Metapopulation models, originally proposed in the field of ecology \cite{hanski,metapop1,metapop2}, enable the mixing of mobility and contagion dynamics into a single formulation. These models can be described as networks in which nodes account for geographic locations (such as neighborhoods in cities, cities within countries, etc), \textit{ i.e.} subpopulations where large collectivities live and interact. In addition, the links of the network represent (and quantify) trips made by individuals between different subpopulations. 

From the first studies making use of metapopulations for the study of infectious diseases transmission  \cite{pion1,pion2}, the field has advanced both in its theoretical grounds \cite{colizza1,colizza2,colizza3,colizza4,rohani1,rohanibook} and its use for large-scale agent-based simulations \cite{eubank,gleam1,gleam2}. The latter approach incorporates many realistic aspects of human interactions with the goal of being useful for making epidemic forecast and the design of efficient prevention policies. On the other hand, the theoretical part has been spurred by the increasing spatio-temporal resolution of current data gathering techniques and many efforts have been devoted to bridge the gap between theory and realistic models during the last years \cite{ball}.  First attempts in this direction involved displacement kernels \cite{ferguson,truscott} to model a local range of movement around an area that, in the last years, lead the way to the inclusion of  more sophisticated mobility patterns, such as the commuting nature of human mobility \cite{commutes,prx,jtb,natphys}, the high order memory of human displacements \cite{jrsijoan}, or the coexistence of different transportation behaviors \cite{prxmulti}. These sophisticated theories allow to capture the temporal and geographical spread of diseases while providing insights about the mechanisms driving the observed patterns.

In the case of VBD the use of metapopulation models has been recently fostered due to recent outbreaks such as Zika and Chikungunya. This way, epidemic models for VBD transmission have abandoned mean-field and well mixed hypothesis to consider patchy environments subjected to human flows. On the theoretical side, and pretty much as for metapopulations models of human-human transmission diseases, the frameworks rely on important assumptions that allow its analytical study. One of these assumptions is to consider the random diffusion of humans across patches \cite{Xiao,Prosper,Auger} or displacement kernels \cite{chao} instead of realistic mobility patterns. On the other hand, when actual ingredients of human mobility, such as its recurrent nature, are taken into account both random diffusion \cite{Cosner,Moulay} and displacement kernels models \cite{wesolowski}  fail in providing insights about the role that real mobility networks play on the transmission of VBD. Thus, metapopulation theories are still far from incorporating the many aspects influencing the onset of VBD outbreaks and lack the predictive power provided by data-driven agent-based simulations \cite{Zhang,Espana}. 

The main goal of this work is to provide a benchmark that allows the study of large-scale vector-borne epidemics in a unified way and, more importantly, enabling the test of coordinated control strategies at the light of available data about vector incidence together with human demography and mobility datasets. To this aim, we first elaborate a metapopulation model for the transmission of VBD that allows us to derive the conditions under which epidemics take place. The analytical expression of the epidemic threshold is revealed by a matrix encoding the probability of crossed infections between humans and vectors of different subpopulations. Importantly, the spectral analysis of this matrix reveals the risk associated to each patch, pointing out those subpopulations triggering the epidemic onset. We confirm these results by testing synthetic metapopulations and a real case, the city of Cali (Colombia). To round off, taking advantage of the theoretical insights and analytical expressions provided by the formalism, we propose a metric capturing the risk associated to each patch. We implement this metric in the city of Cali obtaining a very good agreement between the estimated risk and the actual distribution of Dengue incidence across districts, highlighting the important role of recurrent human mobility patterns for explaining the spatial dissemination of VBD. 

\section*{Results}
\subsection*{Metapopulation model for VBD transmission}

In the following, we will focus on the description of a vector-borne contagion dynamics in a complex metapopulation. To this aim, we consider a set of $N$ populations or patches in which contagion processes occur. In particular, we consider that the dynamic inside each patch is governed by the Ross-Macdonald (RM) model, as it captures the essential ingredients involved in VBD that do not confer immunity to reinfections such as Dengue. Other models including different disease-specific contagion patterns, such as human-human contagions in Zika \cite{scarpino1,scarpino2}, can be easily accommodated in the introduced framework. The relevant variables of the RM dynamics are: {\em (i)} the fraction of infected humans at time $t$, $\humI(t)$, and {\em (ii)} the fraction of vectors infected at time $t$, $\mosI(t)$. The evolution of these two variables is given as a product of the elementary processes described in Fig. \ref{fig1}a. Namely, susceptible humans become infected with probability $\lambda^{MH}$ after being bitten by an infected vector whereas healthy vectors become infectious with probability $\lambda^{MH}$ when interacting with an infected human. In addition, we assume that each vector makes a number of $\beta$ contacts with (healthy or infected) humans. This way, no human-human or vector-vector direct infections are allowed.  Finally, infected humans become susceptible with probability $\mu^{H}$, while (healthy or infected) vectors die with probability $\mu^{M}$, being replaced by newborn healthy ones. 

Although the RM dynamics captures the elementary contagion processes taking place inside each population, the dynamical evolution of each patch depends strongly on the others, since they are not isolated. On the contrary, many individuals with residence in one subpopulation may visit others during, for instance, their daily commutes to other geographical locations. 
On the other hand,  the mobility of vectors is rather limited, hence they are assumed not to move from their original population. This assumption is valid for many VBD such as Dengue, Zika or Chikungunya since their carriers, {\em Aedes} mosquitoes, typically fly an average of 400 meters \cite{dispersal}. Thus, it is the mobility of infectious individuals (who pass the diseases to healthy vectors living in distant subpopulations) what triggers the propagation of local disease outbreaks across the whole system.
\begin{figure*}[t!]
\centering
\includegraphics[width=\textwidth]{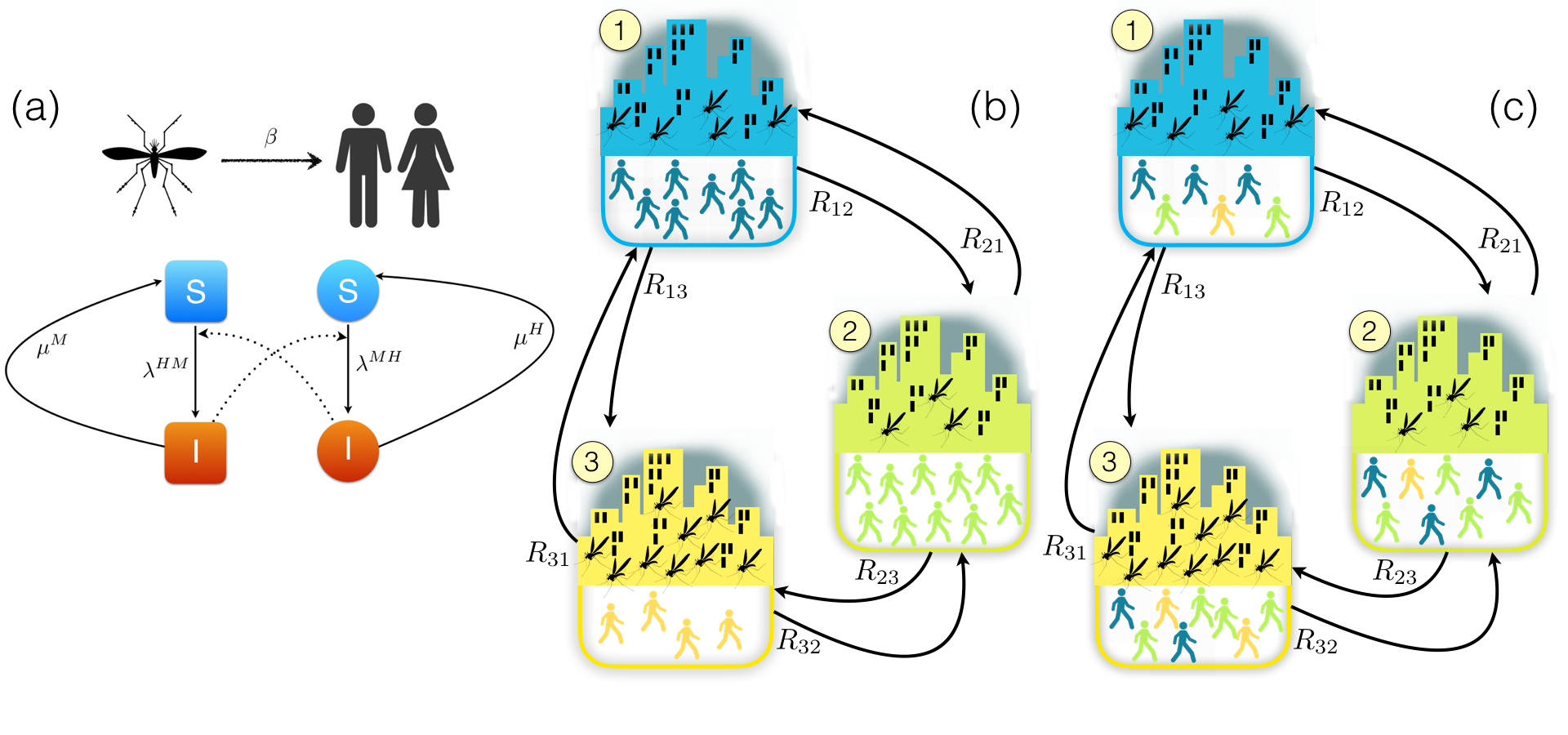}
\caption{{\bf Ross-Macdonald model and metapopulation approach}. (a) Schematic representation of the processes described in the RM model. Relevant parameters are: {\em (i)} the probability that an infected vector transmits the disease to a healthy individual, $\lambda^{MH}$,  {\em (ii)} the probability that an infected human transmits the disease to a healthy vector, $\lambda^{HM}$, {\em (iii)} the feeding rate of vectors $\beta$, {\em (iv)} the probability that an infected human recovers, $\mu^{H}$, and {\em (v)} the mortality rate of vectors, $\mu^{H}$. In (b) and (c) we show schematically the two basic stages of each Monte Carlo step in our metapopulation approach. As shown in (b), individuals are associated to one of the 3 nodes of a network.  Namely, starting from the subpopulation in the top ($1$) and following clockwise we have populations composed of 8, 10 and 4 humans with 5, 3 and 9 vectors respectively. Once the movement has been done [see panel (c)] individuals mix and, consequently, the instant populations of humans at each node change to 6, 8 and 8 . The RM dynamics then takes place among the individuals and vectors coexisting at that moment in the same subpopulation. Finally, after these interactions, the individuals go back to their respective associated nodes and the configuration is again the one of panel (b). Note that we consider that vectors do not move from their corresponding node.}
\label{fig1}
\end{figure*}
To characterize the mobility of individuals we denote each population as a node of a graph. Each node $i$ has a population of $n_i$ individuals and $m_i$ vectors and, importantly, they may be different from one population to the other, as  they are derived from the demographic partition of the population and the observed vector prevalence in each patch. It is important to recall that each individual is associated to one subpopulation, say $i$, considered as her residence. This way, the population $n_i$ of the patch $i$ is the number of individuals whose residence is node $i$. In its turn, nodes are connected in pairs forming a complex weighted and directed network encoded in an adjacency matrix ${\bf R}$, whose entries $R_{ij}$ account for the  the probability that a trip departing from patch $i$ has as destination population $j$ (see Figure \ref{fig1}b). Matrix ${\bf R}$ can be computed from the observed number of trips between each pair of nodes ($i$, $j$), $W_{ij}$, as:
\begin{equation}
R_{ij}=\frac{W_{ij}}{\sum_{j=1}^{N}W_{ij}}\;.
\label{eq:R}
\end{equation}
Thus, matrix ${\bf R}$ encodes the information provided by mobility datasets . 

The former two dynamical processes at work --RM dynamics and human movements-- interplay in each time step of the metapopulation dynamics as follows. We start with a small fraction of infected humans and/or vectors. The initial quantity, being small, can be homogeneously distributed across the populations or localized in one or few nodes in case of being interested in determining those patches boosting epidemic spreading in its early stage. Once the initial infectious seed has been placed, the simulation steps per unit time are the following:
\begin{itemize}
\item At each time step, $t$, healthy agents decide to move from their residence with probability $p$ or remaining in it with probability $(1-p)$. Moreover, as symptoms associated to some VBD are severe, we also include the possibility of re-scaling the infected agents' mobility to $\alpha p$ with $\alpha \in [0,1]$.
\item If an agent leaves her residence, say $i$, she goes to a different subpopulation chosen among those connected to $i$. The choice is dictated by matrix ${\bf R}$ in Eq. (\ref{eq:R}), being $R_{ij}$ the probability of moving from $i$ to subpopulation $j$.
\item Once all the individuals have been placed in the patches (see Fig.~\ref{fig1}.c), humans and vectors that are currently at the same patch interact as dictated by the RM model. This way, both humans and vector update their dynamical states (Susceptible or Infected as shown in Fig. \ref{fig1}.a) as a result of the contagion and recovery processes.
\item Once the epidemic state of the agents have been updated, each individual moves back to her residence and the process starts again for time $t+1$.
\end{itemize}

\subsubsection*{Markovian Formulation} 

Once defined the basic steps of the mechanistic simulations, we now tackle the mathematical formulation of the processes described above. To this aim, for each patch $i$ ($i=1,...,N$) we have $2$ variables: the probabilities that humans with residence in $i$, $\rho^{H}_i(t)$, and vectors associated to $i$, $\rho_{i}^{M}(t)$, are infectious at time $t$ respectively. These $2N$ variables evolve according to the following Markovian equations:
\begin{eqnarray}
\rho^{H}_{i}(t+1) & = & \rho^{H}_{i}(t)(1-\mu^H) + (1-\rho^{H}_{i}(t))I^{H}_i(t),\label{infected_h}\label{Markovian1}\\
\rho^{M}_{i}(t+1) & = & \rho^{M}_{i}(t)(1-\mu^M) + (1-\rho^{M}_{i}(t))I^{M}_i(t),
\label{Markovian2}
\end{eqnarray}
where $I^{H}_i(t)$  and $I^{M}_i(t)$ account for the probability that a healthy human with residence in subpopulation $i$ and a healthy vector associated to $i$ are infected at time $t$ respectively. The former infection probability reads:
\begin{equation}
I^{H}_i(t)=(1-p)P_{i}^{H}(t)+p\sum_{j=1}^{N}R_{ij}P_j^{H}(t)\;,
\label{I_h}
\end{equation}
where $P_{i}^{H}(t)$ is the probability that an agent placed in population $i$ at time $t$ is infected. This probability can be written as:
\begin{equation}
P_{i}^{H}(t)=1-\left(1-\lh\rho_{i}^{M}\frac{1}{n_{i}^{eff}(\rho_H(t),\alpha,p)})\right)^{\beta m_i}\;.
\label{P}
\end{equation}
Finally, $n_{i}^{eff}(\rho_h(t),\alpha,p)$, which is the number of humans placed in (but not necessarily resident in) population $i$ can be expressed as:
\begin{eqnarray}
n_{i}^{eff}(\rho_H(t),\alpha,p)&=&\left[1-p\left(1-(1-\alpha)\rho_i^H(t)\right)\right]n_{i}\nonumber \\
&+& p\sum_{j=1}^{n}R_{ji}\left(1-(1-\alpha)\rho_j^H(t)\right)n_{j}\;.
\label{eq:neff}
\end{eqnarray}

In the same fashion, the expression for  $I_{i}^{M}(t)$ in Eq. (\ref{Markovian2}) reads:
\begin{equation}
I^{M}_i(t)=1-\left(1-\lambda^{HM}\frac{i^{eff}_{i}(t)}{n_{i}^{eff}}\right)^{\beta}
\end{equation}
where $i^{eff}_{i}(t)$ is the number of infected humans placed in population $i$ at time t:
\begin{equation}
i^{eff}_{i}(t)= (1-\alpha p)n_i\rho^{H}_{i}(t)+\alpha p\sum_{j=1}^{N}R_{ji}n_j\rho^{H}_j(t)\;.
\label{eq:ieff}
\end{equation}

The above equations describe the time evolution for the VBD incidence, $\vec{\rho}^{H}(t)=\{\rho^{H}_i(t)\}$ and $\vec{\rho}^{M}(t)=\{\rho^{M}_i(t)\}$, in a collection of connected patches with arbitrary demographic, $\vec{n}$, and vector, $\vec{m}$, distribution. They enable, by integrating the equations starting from a given initial condition $\vec{\rho}^{H}(0)$ and $\vec{\rho}^{M}(0)$, to monitor the geographical spread of VBD and to analyze the stationary incidence of the diseases across patches, while the global incidence of the disease can be calculated as the fraction of infected humans in the entire system $\rho^H=\frac{\sum_i \rho_i^H n_i}{\sum_i n_i}$. The accuracy of Eqs. (\ref{Markovian1})-(\ref{Markovian2}) has been tested with the results derived from mechanistic simulations in synthetic metapopulations (see Methods). As shown in Figures S3 and S4 of the SI, the perfect agreement found both for stationary solutions and the spatio-temporal evolution points out the validity of the Markovian formulation.

\begin{figure*}[t!]
\centering
\includegraphics[width=5.5in]{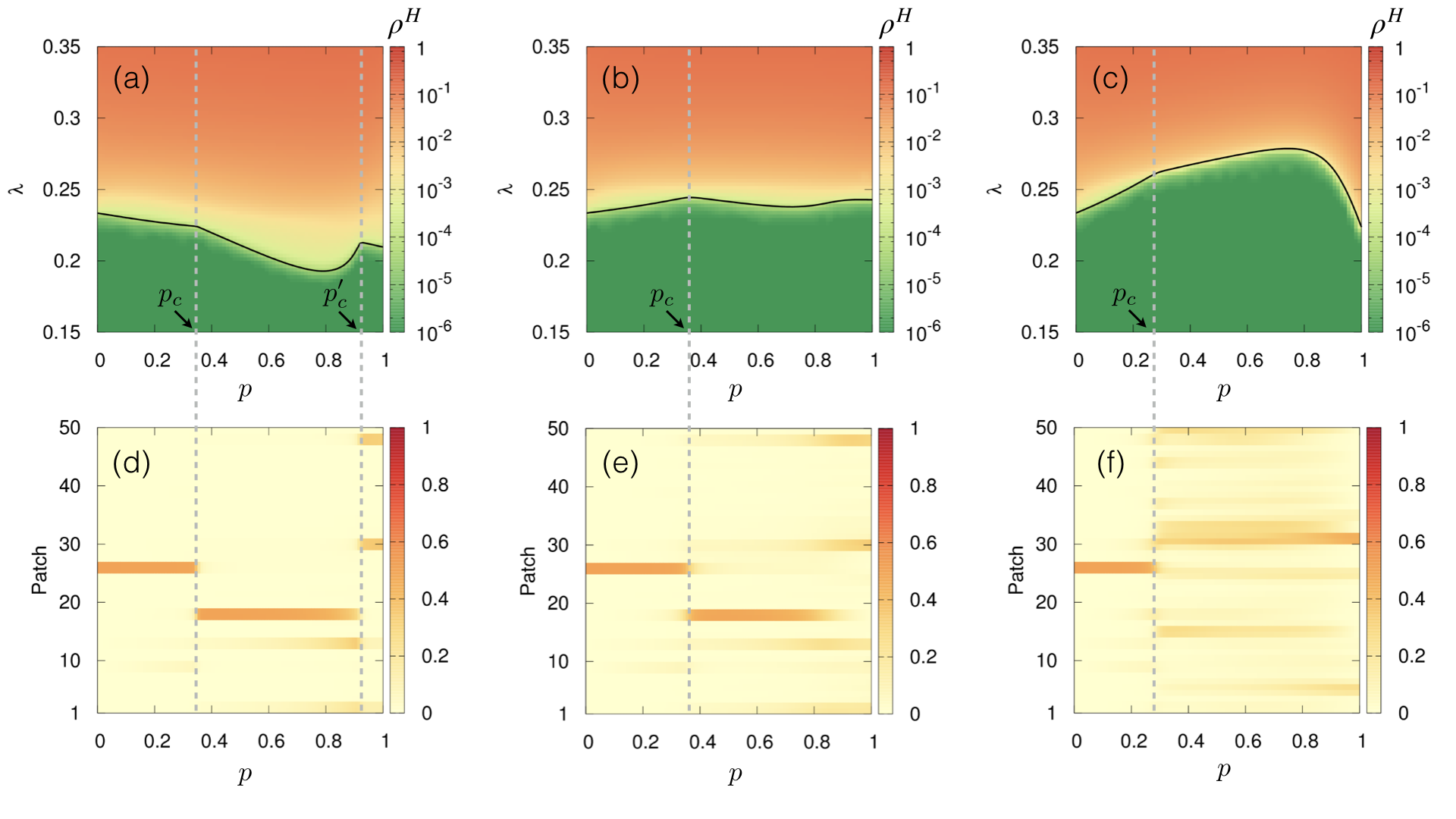}
\caption{{\bf Epidemic threshold and evolution of the leading eigenvector of matrix ${\bf M\widetilde{M}}$}. Panels (a)-(c) in the top show three epidemic diagrams $\rho^{H}(p,\lambda)$ in a synthetic metapopulation of $N=50$ patches.
Each panel corresponds to a different re-scaling value ($\alpha$) for the mobility of infected humans, namely: (a) $\alpha=0$, (b) $\alpha=0.5$ and (c) $\alpha=1$. In addition, we have set $\lambda^{HM}=\lambda^{MH}=\lambda$ while the rest of the RM parameters are: $\mu^{H}=0.3$, $\mu^{M}=0.3$, and $\beta=1.0$. The color code show the incidence $\rho^{H}$ as obtained from agent-based simulations while the solid curve represent the prediction for the epidemic threshold, $\lambda_c$, calculated from Eq. (\ref{threshold}). The bottom panels (d)-(f) show the evolution, as a function of $p$, of the $N$ components of the eigenvector of matrix ${\bf M\widetilde{M}}$ corresponding to maximum eigenvalue $\Lambda_{max}({\bf M\widetilde{M}})$.}
\label{Fig2}
\end{figure*}

\subsubsection*{Estimation of the epidemic threshold}
The validation of Eqs. (\ref{Markovian1})-(\ref{Markovian2}) offers the possibility of saving computational costs by integrating the $2\times N$ Markovian equations instead of performing lengthy agent-based simulations. However, the advance behind Eqs. (\ref{Markovian1})-(\ref{Markovian2})  is that they also allow for deriving metrics and analytical results about the dynamical behavior of VBD in large metapopulations. A relevant quantity that can be analyzed is the epidemic threshold, {\it i.e.}, those conditions that turns the epidemic state into a stable solution of Eqs. (\ref{Markovian1})-(\ref{Markovian2}). To address it, we suppose that the infection probabilities of humans and vectors in the stationary state are very small: $\rho_i^H=\epsilon_i^H \ll 1\, ,\rho_i^M=\epsilon_i^M \ll 1\,\forall i$.  This assumption allows us to linearize Eqs. (\ref{Markovian1})-(\ref{Markovian2}) and, after some algebra (see SI), the stationary solutions for the infection probabilities of humans, $\vec{\epsilon}^{H}$ read: 
\begin{equation}
\vec{\epsilon}^H = \frac{\beta^2\lambda^{MH}\lambda^{HM}}{\mu_M\mu_H}\left({\bf M\widetilde{M}}\right)\vec{\epsilon}^H \ ,
\label{eq:eigenvalue}
\end{equation}
where the entries of matrices ${\bf M}$ and ${\bf \widetilde{M}}$ (see SI for their derivation) take the following form:
\begin{eqnarray}
M_{ij}=\left( pR_{ij}\frac{m_j}{\tilde{n}_j^{eff}}+(1-p)\delta_{ij}\frac{m_i}{\tilde{n}_i^{eff}} \right)\label{eqMhum}\;,\\
\widetilde{M}_{ij}=\left(\alpha pR_{ji}\frac{n_j}{\tilde{n}_i^{eff}}+(1-\alpha p)\delta_{ij}\frac{n_i}{\tilde{n}_i^{eff}} \right)\;\;,
\label{eqMmosq}
\end{eqnarray}
where $\tilde{n}_i^{eff}$ is defined as $\tilde{n}_i^{eff}=n_i^{eff}(0,\alpha,p)$. Note that the form of the elements of these two matrices depends on both the mobility properties ($p$, ${\bf R}$) and the demographic distribution of both agents and vectors ($\vec{n}$, $\vec{m}$).

From Eq, (\ref{eq:eigenvalue}) it is clear that nontrivial solutions for ${\vec{\epsilon}^H}$ correspond to the eigenvectors of matrix ${\bf M\widetilde{M}}$. Specifically, given a metapopulation defined by $\vec{n}$, $\vec{m}$, ${\bf R}$ and $p$, the stationary solutions with infinitesimal incidence correspond to eigenvectors of ${\bf M\widetilde{M}}$ whose eigenvalues fulfil:
\begin{equation}
\Lambda_i =  \frac{\mu^M\mu^H}{\beta^2\lh\lm}\; ,
\end{equation}
Under these conditions, the maximum eigenvalue $\Lambda_{max} ({\bf M\widetilde{M}})$ encodes the combination of the RM parameters that corresponds to the epidemic threshold, namely: 
\begin{equation}
 \frac{\beta^2\lh\lm}{\mu^M\mu^H}\Lambda_{max} = 1\ .
\end{equation}
The former equation reveals the minimum infectivities, either $\lambda^{HM}$ or $\lambda^{MH}$, that trigger the epidemic outbreak. To derive a simple critical infectivity one can set $\lh=\delta\lm$, so that:
\begin{equation}
\lh_c= \sqrt{\frac{\mu_H\mu_M}{\delta\beta^2\Lambda_{max}({\bf M\widetilde{M}})}}
\label{threshold}
\end{equation}

To test the validity of Eq. (\ref{threshold}) we have carried out extensive numerical simulations in synthetic metapopulations (see the SI for a detailed description) considering  $\lh=\lm=\lambda$. The top panels in Fig.~\ref{Fig2} show the epidemic diagrams $\rho^H(p, \lambda)$ by computing the fractions of humans infected, $\rho^{H}$, as a function of $\lambda$ and $p$. From these diagrams it becomes clear that, for each value of $p$, there exist a critical value $\lambda_c$ so that for $\lambda>\lambda_c$ the epidemic phase appears. The border of this region (solid curves in Fig.~\ref{Fig2}.a-c)) is the function $\lambda_c(p)$ calculated with Eq. (\ref{threshold}), showing an excellent agreement with the results from numerical simulations.

\subsection*{Abrupt transitions of leading pacthes}

Apart from the agreement between Eq. (\ref{threshold}) and the numerical simulations, the evolution of the epidemic threshold, $\lambda_c(p)$, reported in the three upper panels points out a non-trivial dependence with the degree of human mobility. Contrary to what naively expected, human mobility can be detrimental to epidemics, as clearly illustrated in the panels for $\alpha=0.5$ and $\alpha=1.0$. This counterintuitive effect of mobility was already found for SIR and SIS diseases in networked metapopulations \cite{natphys} as a result of the redistribution of the effective populations across patches due to mobility. In the case of VBD, this process corresponds to an homogeneization of the effective ratios between vectors and humans so that a high risk patch with large $\gamma_i=m_i/n_i$ tends to decrease its effective value  due to the increase of $n_i^{eff}$, Eq.(\ref{eq:neff}), caused by the mobility. 

A more striking phenomenon reported in the epidemic diagrams of Fig.~\ref{Fig2} is revealed by the sharp variations in the slope of the curves $\lambda_c(p)$. These abrupt changes are the product of collisions between the two maximum eigenvalues of matrix ${\bf M\widetilde{M}}$ as $p$ varies. This way, the two maximum eigenvalues interchange their order at some critical mobility value $p_c$. These collisions do not have an strong impact in the epidemic threshold since the function $\lambda_c(p)$ is continuous. However, they are the fingerprint of a sudden change in the form of the eigenvector corresponding to the maximum eigenvalue, $\vec{v}_{max}$, of matrix ${\bf M\widetilde{M}}$. This abrupt transition is of utmost importance since the components of $\vec{v}_{max}$ encode the most important patches driving the unfolding of the epidemics.

Let us recall that matrix  ${\bf M\widetilde{M}}$ incorporates the demographic information, $\vec{n}$, the mobility patterns, $\bf{R}$ and the vector distribution, $\vec{m}$, having as unique parameter the degree of mobility $p$. Thus, for each value of $p$ the spectral analysis of  ${\bf M\widetilde{M}}$ gives us the epidemic threshold $\lambda_c$ and the distribution of patches triggering the epidemic onset in the components of $\vec{v}_{max}$. The evolution of the components of $\vec{v}_{max}$ as a function of $p$ is shown in the bottom panels, (d-f), of Fig.~\ref{Fig2}. 
From these plots it becomes clear that the discontinuities of the slope of $\lambda_c(p)$ correspond to abrupt changes in the form of $\vec{v}_{max}$.  Namely, in the three cases patch number $26$ is the one causing the epidemic onset for $p=0$ and $p\ll1$. This is obvious since patch $26$ is the one with largest ratio $\gamma_i=m_i/n_i$ in the synthetic metapopulation. However, as $p$ increases, the leading patch changes, being replaced by patch $18$ in the case of $\alpha=0$ (d) and $\alpha=0.5$ (e) while for $\alpha=1$ (f) the leading patch is replaced by a collection of them. Remarkably, the case  $\alpha=0$ shows a second abrupt transition at $p'_c\simeq 0.92$. These abrupt changes point out that containment strategies targeting a certain neighborhood can change sharply from efficient to useless due to changes in human mobility, as confirmed in the SI.
%

\subsection*{Assessing the epidemic risk of patches with ${\bf M\widetilde{M}}$}

Spurred by the ability of the Markovian formalism to capture the dynamics of VBD and the insights provided by the spectral properties of matrix ${\bf M\widetilde{M}}$ for identifying the areas triggering the onset of epidemics, we move one step further and evaluate the epidemic risk associated to each patch. To this aim, we propose a theory-driven prevalence indicator which serves as a proxy to determine the most exposed areas to the spread of VBD. 

For this purpose, let us analyze the elements of the matrices ${\bf M}$ and ${\bf{\widetilde{M}}}$, defined in Eqs.~(\ref{eqMhum})-(\ref{eqMmosq}). From these equations we realize that the elements $M_{ij}$  ($\widetilde{M_{ij}}$) contain all the possible microscopic contagion processes from vectors (humans) associated to patch $j$ to humans (vectors) associated to $i$. Therefore, it is possible to estimate the effective number of human-human contagions mediated by vectors that an individual from subpopulation $i$ receives from those with residence in $j$.  This quantity, denoted as $C_{ij}$, can be obtained taking into account all the possible infection pathways connecting humans in patch $j$ to those of $i$  as described in Eq.~\ref{eq:C}. Those infections may take place in three possible ways: {\it(i)} an infected individual from patch $j$ visits patch $i$ and infect a vector that will  later pass the disease to a resident of patch $i$; {\it (ii)} a resident of patch $j$ infects a vector  and a healthy human traveling to $j$ from $i$ gets  infected; {\it (iii)} both the infected individual from $j$  and the healthy individual from $i$ travel to a contiguous third patch $k$ where the infection takes place mediated by a vector.

\begin{figure}[b!]
\centering
\includegraphics[width=4in]{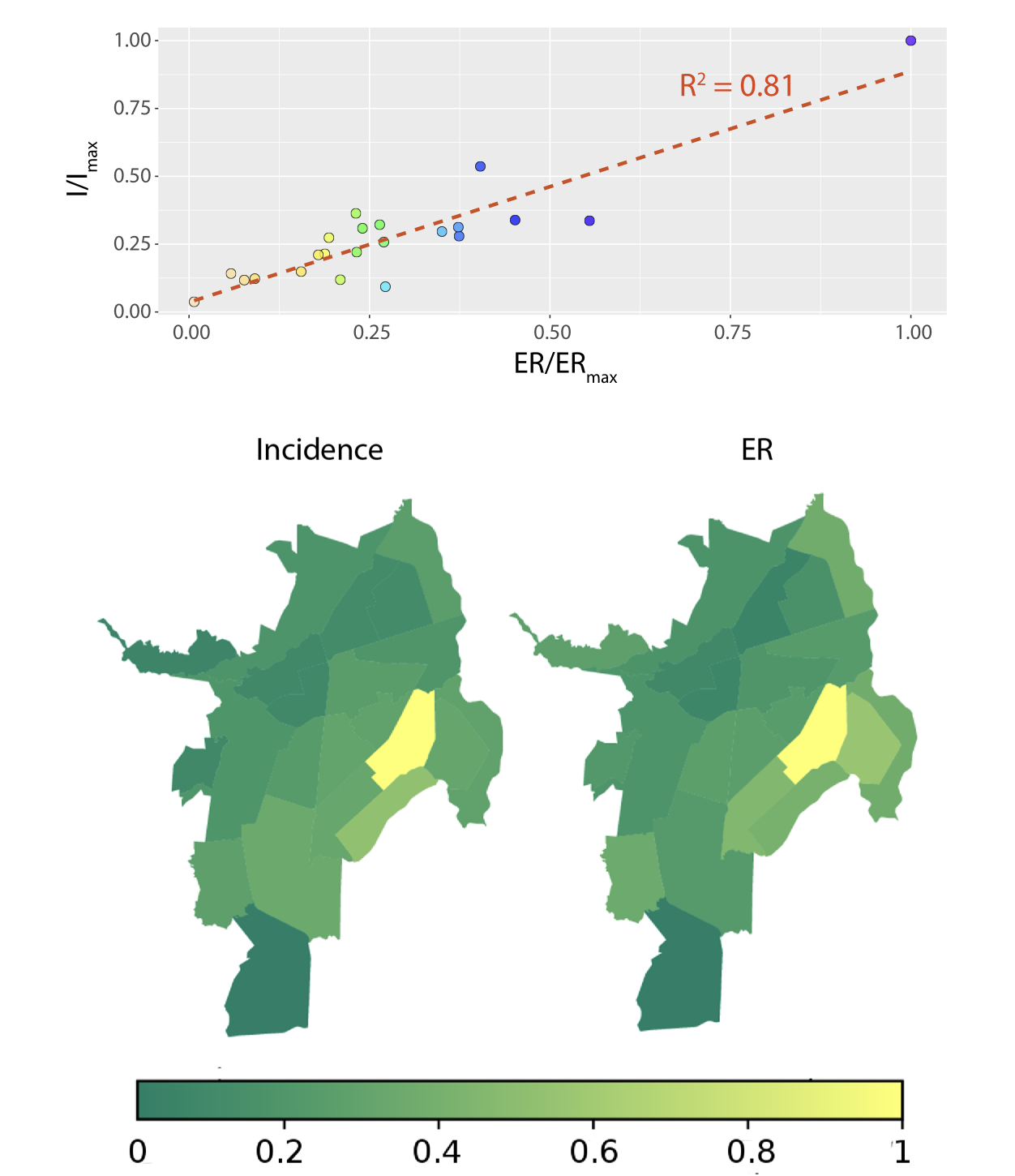}
\caption{{\bf Real Dengue incidence {\it versus} estimated epidemic risk in the city of Cali (Colombia).} Top: Normalized epidemic risk ($ER/ER_{max}$) {\it versus} normalized Dengue incidence ($I/I_{max}$) for each of the $22$ districts of Cali. Color encodes the Epidemic Risk, from the lower (yellow) to the highest (blue).The correlation between the two variables yields a coefficient of determination of $R^2=0.81$. Bottom: Spatial distributions of the normalized Dengue incidence in the city of Cali (left) and the normalized epidemic risk (right) according to Eq. (\ref{eq:ER}). The parameters concerning agents' mobility have been set to $(p,\alpha)=(0.36,0.75)$ (see SI for more details).}
\label{fig3}
\end{figure}
\begin{equation}
C_{ij} = \sum\limits_{k=1}^N M_{ik}\widetilde{M_{kj}}\;.
\label{eq:C}
\end{equation}
Finally,  to make predictions about the impact of the disease on a geographical area, $i$, we must account for all the possible infections from each patch of the metapopulation and to weight the resulting number by the population of $i$. This way, the epidemic risk indicator for each patch $i$, in the following denoted as $ER_i$, can be defined as:
\begin{equation}
ER_i = n_i \sum\limits_{j=1}^N C_{ij} = n_i\sum\limits_{j=1}^N\left({\bf M\widetilde{M}}\right)_{ij}\;.
\label{eq:ER}
\end{equation}

The evaluation of $ER_i$, as defined above, can be done directly without the need of neither numerical simulations nor making the integration of the Markovian equations. In fact, once  data about demography, vector distribution and mobility patterns are available, one can estimate the epidemic risk of each subpopulation. In the SI we show the validation of the risk indicator for the synthetic metapopulations  by comparing the values $ER_i$ with the disease incidence as obtained from the mechanistic simulations. The fair agreement (coefficient of determination $R^2=0.95$) between our indicator and the results from numerical simulations allows us to rank the patches according to their exposure to VBD without the need of performing computationally expensive simulations or integrating an large set of Markovian equations.

\subsubsection*{A real metapopulation: Santiago de Cali (Colombia)} 

To validate further the Epidemic Risk measure we now move to a real metapopulation, the city of Santiago de Cali (Colombia). With a population of more than 2 millions of inhabitants, it offers the possibility of comparing our predictions in a scenario for which severe epidemic outbreaks of VBD are recurrently found. In particular, due to its location and climate, Cali is a Dengue endemic area in which records of the historical incidence of this disease are available for comparison. To this aim, we collected demographic and mobility datasets \cite{mobility} whereas vectors abundance across districts was obtained from entomological reports made yearly by the local authorities \cite{vectors}. 

With this information at hand, and using Eq.~(\ref{eq:ER}), we assign the epidemic risk of each of the $22$ districts in which the city is divided. These values are compared to the observed Dengue incidence across the $22$ patches during the period 2015-2016 \cite{incidence} (details in the SI). In Fig. (\ref{fig3}) we show this comparison by normalizing the values of both epidemic risk and Dengue incidence by their maximum observed value (in both cases that of district 13). In particular, we find a coefficient of determination of $R^2=0.81$, indicating that  the proposed prevalence indicator is able to capture the spatial distribution of Dengue cases across the city. On more general grounds, this agreement points out that given the demography, the commuting patterns and the spatial distribution of vectors across a given population, one can use Eq.~(\ref{eq:ER}) to identify areas where containment measures should be promoted to reduce the impact of possible outbreaks. The strong dependence of $R^2$ on human mobility (see sensitivity analysis in the SI) points out the prominent role that human mobility plays in the dissemination of VBD and, therefore, its relevance for the design of efficient policies to prevent local outbreaks from spreading across populations.


\section*{Discussion}

The control of infectious diseases represents one of the major societal challenges. Understanding the complex interdependency between human activity and contagion processes is key to explain the onset and development of large-scale epidemics. Here, focusing on VBD, we have integrated information from urban daily commutes and the geographic distribution of humans and vectors to estimate the epidemic risk associated to different connected regions. In particular, we have provided a metapopulation formalism to assess the role that the former ingredients play on the propagation of VBD. We have proved that this formalism constitutes a very reliable and time-saving platform, since its Markovian equations enable to reproduce very accurately not only the global incidence of VBD but also the spatio-temporal spreading patterns observed in Monte Carlo simulations. 

Based on this agreement, we have derived an analytical expression of the epidemic threshold that captures the critical conditions which leads to the onset of epidemics. Apart from the detrimental effect that mobility may have on the spread of diseases, the study of the epidemic threshold has revealed interesting phenomena such the existence of abrupt changes in the way epidemics unfold. In particular, we have shown that the subset of patches leading the epidemic onset can suddenly change as human mobility varies. This phenomenon highlights the need of incorporating real human mobility patterns into the design of containment policies targeting specific geographical areas, for efficient policies can turn useless due to a small variation of human mobility habits. 

Finally, relying on the matrix containing the information about the effective number of human-human contagions, we have derived an epidemic risk indicator that allows us to classify the patches according to their exposure to VBD. By computing this epidemic indicator, we have reproduced with great accuracy the geographical distribution of Dengue incidence in the city of Santiago de Cali (Colombia), where Dengue in an endemic disease, thus being able to identify the most vulnerable areas where prevention measures should be promoted.

In a nutshell, our results point out that the spread of VBD is the result of a delicate interplay between commuting flows, human census and vector distribution. This interplay is captured both in the analytical expression of the epidemic threshold and in the epidemic risk indicator. As a result of it, we have shown that small variations of the former ingredients, such as the degree of mobility, can lead to abrupt changes in the way epidemics unfold. Our framework, although containing several simplifying assumptions to allow the analytical treatment, has shown useful to integrate human and contagion dynamics and it can be readily implemented to identify those regions where immunization policies should be reinforced and to forecast the consequences of control strategies focused on mobility restrictions.


\section*{Methods}
\subsection*{Mechanistic numerical simulations}
With the aim of assessing the accuracy of the Markovian formulation we compared the predictions obtained using  Eqs.~\ref{Markovian1}-\ref{eq:ieff} --for both the epidemic incidence and the spatio-temporal evolution of the disease-- with numerical results from extensive mechanistic simulations. 

In the simulations with synthetic networks we start by setting the human population of each node ($n_i$) to a constant value ($n_i=10^3$), while vectors population $m_i$ is set to be proportional to $n_i$ as  ($m_i =\gamma_i n_i$) with $\gamma_i$ varying from one population to another and extracted from a uniform distribution within the range $\gamma_i \in [0.3,1.7]$.  For the mobility network of the city of Cali instead, we set $n_i$ and $m_i$ according to census and mosquitos proxy data publicly available from the municipality of Cali (see the SI for details). All the populations are initially composed only by healthy individuals until a seed of the disease is introduced in the system.  For the seed  we consider two different options: one in which the $1\%$ of the agents in each sub-population is initially infected and a second one where the seed is localized in a single sub-population. 

After the seed is introduced, the reactive and diffusive processes take place with the same time-scale. At each time step, agents decide whether to move to a neighboring patch or to remain in their home patch. To do so, each agent generates an independent and identically distributed random variable (i.i.d.) $r$ between [0,1]. If $r$ is smaller than the moving probability $p$ the agent will move otherwise, she will stay in her node. In addition, to model the possible impairment produced by the disease, we also assume that infected individuals could be less prone to move by rescaling their mobility to $\alpha p$, with $0 \le \alpha \le 1$. In any case, if the agents moves, another i.i.d. $r'$ determines the destination patch. $r'$ is extracted between [0,$\sum_j W_{ij}$], where $W_{ij}$ is the weight of the link between populations $i$ and $j$. Then, the destination is chosen as the first node $k$ that satisfies $\sum_k W_{ik}\geq r'$, assuring that the destination is selected proportionally to the mobility flux between sub-populations $i$ and $k$.

Once the agents reached their new location, an iteration of the contagion and recovery process starts. To model the infection dynamics of VBD inside each patch we rely on the Ross-Macdonald model on well mixed populations. Inside each population, each vector, regardless of its state, bites $\beta$ randomly selected individuals. In the case that the vector is infected and the human is healthy the disease will spread with probability $\lambda_{MH}$. In the opposite case, the vector gets the disease with probability $\lambda_{HM}$. Finally, if the vector and the human are both healthy or both infected, nothing happens. Regarding the recovery phase, each infected human recovers with probability $\mu_{H}$ while mosquitos do not recover but are replaced by healthy new individuals at rate $\mu_{M}$.

Simulations finish when the epidemic reaches a stationary state: i.e. or the disease dies out or the fluctuations in the total number of infected humans in the system $\rho^{H}(t)$ are lower than $10^{-5}$ during the last $100$ time steps. Results are then averaged over the different realizations and compared with the theoretical predictions given by the Markovian model.

\subsection*{Synthetic Metapopulations} Synthetic metapopulations used in Fig.~2 are composed of $50$ patches that correspond to the nodes of a network constructed following the preferential attachment model of Barab\'asi and Albert \cite{SF} with mean degree $\langle k\rangle=4$. The elements of matrix ${\bf R}$ are: $R_{ij}=1/k_i$, being $k_i$ the degree of patch $i$. The human population of nodes are identical $n_i=10^3$ while vector populations are randomly assigned in the range $m_i\in[300,1700]$.


\section*{Contributions}
All of the authors designed  research. D.S.-P., J.H.A.-C. ,S.M. and J.G.-G. conducted  research.  J.H.A.-C. and H.J.M. provided and analyzed data. D.S.-P., S.M. and J.G.-G. wrote the manuscript. All the authors revised and approved  the manuscript.


\section*{Competing financial interests}

The authors declare no competing financial interests.

\section*{Acknowledgements}

We are especially grateful to A. Arenas and O. Vasilieva for useful comments and discussions. This work was partially supported by Universidad del Valle (grant  CI-165), Gobierno de Aragon/Fondo Social Europeo (Grant E36-17R),  Ministerio de Economia, Industria y Competitividad (MINECO) and Fondo Europeo de Desarrollo Regional (FEDER) (Grants FIS2015-71582-C2 and FIS2017-87519-P) and by the Agencia Estatal de Investigacion (AEI, Spain) and Fondo Europeo de Desarrollo Regional under Project PACSS Project No. RTI2018-093732-B-C22  (MCIU, AEI/FEDER,UE) and through the Mar\'ia de Maeztu Program for units of Excellence in R\&D (MDM-2017-0711 to IFISC Institute).

\bibliographystyle{naturemag}
\let\oldaddcontentsline\addcontentsline
\renewcommand{\addcontentsline}[3]{}

\let\addcontentsline\oldaddcontentsline

\newpage
\setcounter{figure}{0}
\setcounter{equation}{0}
\renewcommand{\thefigure}{S\arabic{figure}}
\renewcommand{\theequation}{S\arabic{equation}}
\renewcommand{\thesection}{S\arabic{section}}

\begin{Large}
\noindent
{\bf \textsf{Supplementary information}}
\end{Large}
\baselineskip24pt
\tableofcontents

\section{Ross-Macdonald model}
The Ross-Macdonald (RM) model\cite{RM1,RM2,RM3}  is one of the most paradigmatic benchmarks for the study of the transmission of vector-borne diseases (VBD). Although many modifications can be made to include ingredients that play a role in the evolution of VBD (such as the seasonal effects that influence the activity of vectors), it captures the essential processes leading to the unfold of a disease in well-mixed populations. In this sense, the RM model has proved to be a very insightful framework for characterizing VBD which do not confer immunity such as Dengue. 

Here, we derive from first principles a slightly modified version of the original RM model which sets the foundation for the metapopulation model. First, we consider a closed population of both vectors and humans which are homogeneously mixed. The model assumes that vectors and humans can be either Susceptible of contracting the disease or Infected. The relevant variables are {\em (i)} the fraction of infected humans at time $t$, $\humI(t)$, and {\em (ii)} the fraction of infected vectors at time $t$, $\mosI(t)$. The evolution of these variables is given by the elementary processes depicted in Fig. \ref{fig1}. Namely, Susceptible humans become Infected with probability $\lambda^{MH}$ after being bitten by an infected vector while healthy vectors get the infection with probability $\lambda^{HM}$ when interacting with an infected human. We also define $\beta$ as the number of contacts each vector has with (healthy or infected) humans, {\em i.e.} the feeding rate of each vector. No direct human-human or vector-vector infections are allowed.  On the other hand,  infected humans and vectors become healthy with probabilities $\mu^{H}$ and $\mu^{M}$, respectively. For humans, this probability is easily understable as the inverse of the infectious period -- the time that an infected requires to overcome the disease--. Considering vectors, this intuition fails since once they get infected they cannot get rid of the pathogen. However, as we are dealing with steady populations, $\mu^{M}$  encodes the renewal of vectors population. All vectors that die are assumed to be replaced by newborn susceptible ones.
 \begin{figure}[t!]
\centering
\includegraphics[width=3.5in]{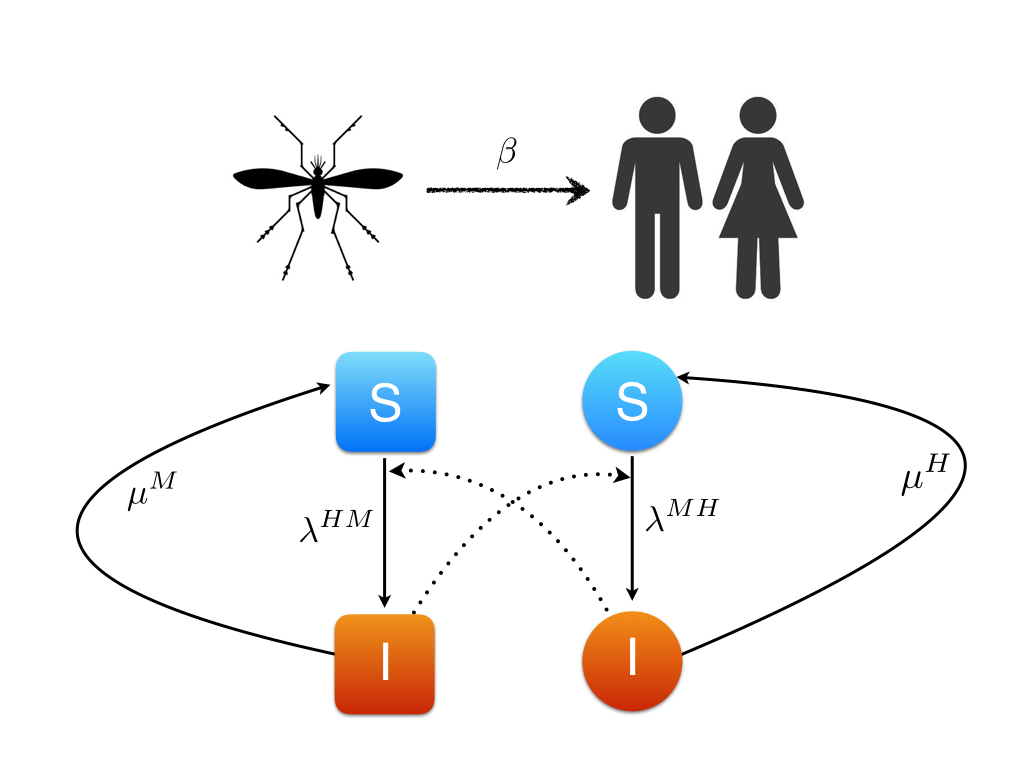}
\caption{Schematic representation of the Ross-Macdonald model. Relevant parameters  are: {\em (i)} the probability that an infected vector transmit the disease to a healthy individual, $\lambda^{MH}$,  {\em (ii)} the probability that an infected human transmit the disease to a healthy vector, $\lambda^{HM}$, {\em (iii)} the feeding rate of a vector $\beta$, {\em (iv)} the probability that an infected human recovers, $\mu^{H}$, and {\em (v)} the mortality rate of vectors, $\mu^{H}$.}
\label{fig1}
\end{figure}
Considering the above processes (see also Fig. \ref{fig1}) we can write the following time discrete dynamical equations for the fractions $\humI(t)$, and $\mosI(t)$ as:
\begin{eqnarray}
\humI(t+1)  &=&  \humI(t)(1-\mu^{H})+(1-\humI(t))I^H(t)\;,
\label{EqH}
\\
\mosI(t+1)  &=&  \mosI(t)(1-\mu^{M})+(1-\mosI(t))I^M(t)\;,
\label{EqM}
\end{eqnarray}
where $N^{H}$ and $N^{M}$ are the number of humans and vectors in the population respectively while $I^{H}(t)$ and $I^{M}(t)$ are the probabilities that a healthy human and a healthy vector get infected at time $t$, and they read as:
\begin{eqnarray}
I^{H}(t)&=&\left[1-\left(1-\lh \mosI(t)\frac{1}{N^H}\right)^{\beta N^{M}}\right]\;,
\label{InfH}
\\
I^{M}(t)&=&\left[1-\left(1-\lm\humI(t)\right)^{\beta}\right]\;.
\label{InfM}
\end{eqnarray}
The first term in the r.h.s. of Eq. (\ref{EqH}) ((\ref{EqM})) describes the fraction of infected humans (vectors) at time $t$ who remains  infected at time step $t+1$ while the second term accounts for the fraction of healthy humans (vectors) that gets infected at time $t+1$. Note also that Eq.(\ref{InfH}) includes an extra term with respect to Eq.(\ref{InfM}), that represents the probability for a human to receive a contact from each vector. We will demonstrate later that this element plays a crucial role on the unfolding of VBD.
\newpage

Equations (\ref{EqH}) and (\ref{EqM}) can be reduced to the classical formulation of the RM model by taking into account that the terms $\lh \mosI(t)$ and $\lm\humI(t)$ are pretty small so that, by using the approximated expression $1-(1-\epsilon)^{x}\simeq x\epsilon $, the continuous-time versions of Eqs. \ref{EqH} and \ref{EqM} turn into:
\begin{eqnarray}
\dot{\rho}^{H}(t) & = & -\mu^H\humI(t)+\beta\lh\gamma(1-\humI(t))\rho^{M}(t),
\label{EqHusual}
\\
\label{EqVusual}
\dot{\rho}^{M}(t) & = & -\mu^M\mosI(t)+\beta\lm(1-\mosI(t))\rho^{H}(t)\;,
\end{eqnarray}
where $\gamma$ is the ratio between the population of humans and vectors, $\gamma=\frac{N^{M}}{N^{H}}$, that can be considered as another parameter of the model.

Equations (\ref{EqHusual}) and (\ref{EqVusual}) represent the usual way the RM model is presented. As shown, they correspond to an approximation that turns to be valid close to the epidemic threshold, {\em i.e.}, when  the fraction of infected vectors and humans are small but nonzero, while the formulation in Eqs.~(\ref{EqH}-\ref{InfM}) is accurate in the entire epidemic diagram. The mathematical analysis of the model yields to an analytical estimation for the epidemic threshold as a function of the parameters $\lh$, $\lm$, $\mu^{H}$, $\mu^{M}$, $\beta$ and $\gamma$. In particular, the first step to compute the threshold is to calculate the nullclines of the system of equations (\ref{EqHusual}) and (\ref{EqVusual}): 
\begin{eqnarray}
\dot{\rho}^{H}&=&0 \;\;\rightarrow\;\;\rho^{M}=A\frac{\rho^{H}}{1-\rho^{H}}
\label{nullH}
\\
\dot{\rho}^{M}&=&0 \;\;\rightarrow\;\;\rho^{M}=\frac{\rho^{H}}{B+\rho^{H}}
\label{nullM}
\end{eqnarray}
where $A=\frac{\mu^{H}}{\beta\lh\gamma}\geq 0$ and $B=\frac{\mu^{M}}{\beta\lm}\geq 0$. The above system is trivially satisfied by the $\rho^{H}=\rho^{M}=0$, the case in which no prevalence of the virus is found in the system. However, we are interested in the stability of the epidemic phase.

\begin{figure}[t!]
\centering
\includegraphics[width=4.0in]{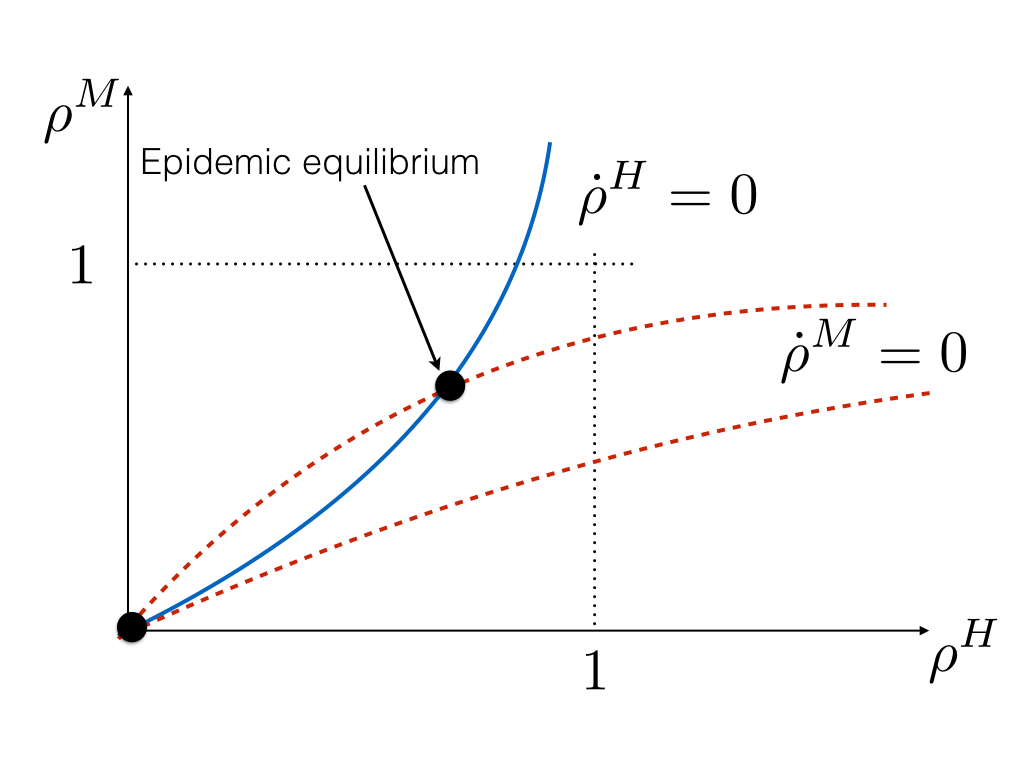}
\caption{Graphical solution of the RM model. The solid (blue) line represents de nullcline of Eq. (\ref{nullH}) whereas the dashed (red) one is that of Eq. (\ref{nullM}). Besides the trivial solution at  $\rho^{H}=\rho^{M}=0$ the epidemic solution exists when the two nullclines intersects at $\rho^{H}>0$.}
\label{fig2}
\end{figure}

From Eq.  (\ref{nullH}) it becomes clear that $\rho^{M}$ diverges as $\rho^{H}\rightarrow 1^{-}$ whereas Eq.(\ref{nullM}) reveal that 
$\rho^{M}$ will continuously grow reaching $\rho^{M}=1$ asymptotically. In Fig. \ref{fig2} we plot the two nullclines (\ref{nullH}) and (\ref{nullM}) with solid (blue) and dashed (red) lines respectively. From the plot, it can be easily seen that the two curves cross for $\rho^{H}>0$ only when the derivative of the dashed nullcline (\ref{nullM}) at $\rho^{H}$ is larger than the one for the solid nullcline (\ref{nullH}). This condition is satisfied whenever
\begin{equation}
A<\frac{1}{B}\;\;\rightarrow\;\;\frac{1}{\gamma\beta^2}\frac{\mu^{H}\mu^{M}}{\lh\lm}<1\ .
\end{equation}
In this sense, the limit of this inequality determines the boundary between the epidemic and disease-free regimes. This boundary is given by the following equation:
\begin{equation}
\frac{\beta^2\lh\lm\gamma}{\mu^H\mu^M}=1
\end{equation}

\section{Description of the Cali dataset}
\label{sec:cali}
One of the most important contribution of this work is the formulation of a new framework which can easily incorporate mobility data of real cities to address real epidemic scenarios. In the main text, we tackle the spread of VBD in the city of Cali (Colombia), whose geographical and meteorological features make it an endemic region for several VBD such as Dengue, Chikungunya or Zika. In particular, here we focus on the spread of Dengue.

To assess the effect of mobility on the spread of Dengue in Cali, it is necessary to reconstruct the mobility network of its inhabitants from data. For this purpose, we divide the city into 22 neighborhoods, which correspond to the official administrative divisions called {\it comunas}. Regarding demography, the population distribution across comunas has been extracted from census data that the municipality facilitates\cite{mobility}. Mobility flows connecting comunas are extracted from urban commuting surveys\cite{surveysmed}. As a result, more than $10^5$ trajectories were recorded, which suppose a representative sample of Cali's commuting flows. Once data have been gathered, an origin-destination matrix, encoded in our formalism by  matrix ${\bf R}$, is computed as:
\begin{equation}
R_{ij} = \frac{W_{ij}}{\sum\limits_{l=1}^N W_{il}}\ ,
\end{equation}
where the numerator corresponds to the number of trips between patches $i$ and $j$  while the denominator counts all the reported trips departing from patch $i$. The result is a weighted directed network encoding the probability that an agent visits other neighborhoods different from its residence. 

Apart from the mobility network, the distribution of vectors across the city also plays a crucial role on the outcome of the disease. The number of vectors inside a geographical region is strongly linked to environmental features such as altitude, temperature and humidity but also to human-dependent factors like health and economic conditions. To model vectors' distribution across patches, we use as a proxy the so-called  {\it recipient index}\cite{Cali2015}. This quantity encodes the probability of finding vector pupae in different recipients which have been previously distributed across the city. A high value of the index means a higher probability of finding vectors. For this reason, we assume that the ratio between the number of vectors and humans inside each patch in our model is directly proportional to its recipient index, which is extracted from a report published in 2014\cite{Cali2015}.

\section{Validation of the Markovian equations}
Here we aim at assessing the ability of our model to predict the impact of VBD in both synthetic and real world metapopulations. To do so, we compare theoretical predictions from our model with results from extensive mechanistic simulations. Theoretical predictions are obtained by iterating Eqs. (2-8) of the main text until a stationary state has been reached, while mechanistic simulations are performed by updating the dynamical state of vectors and humans following the dynamical rules defined in the main text (see {\em Methods}).

\begin{figure}[t!]
\centering
\includegraphics[width=0.86\columnwidth]{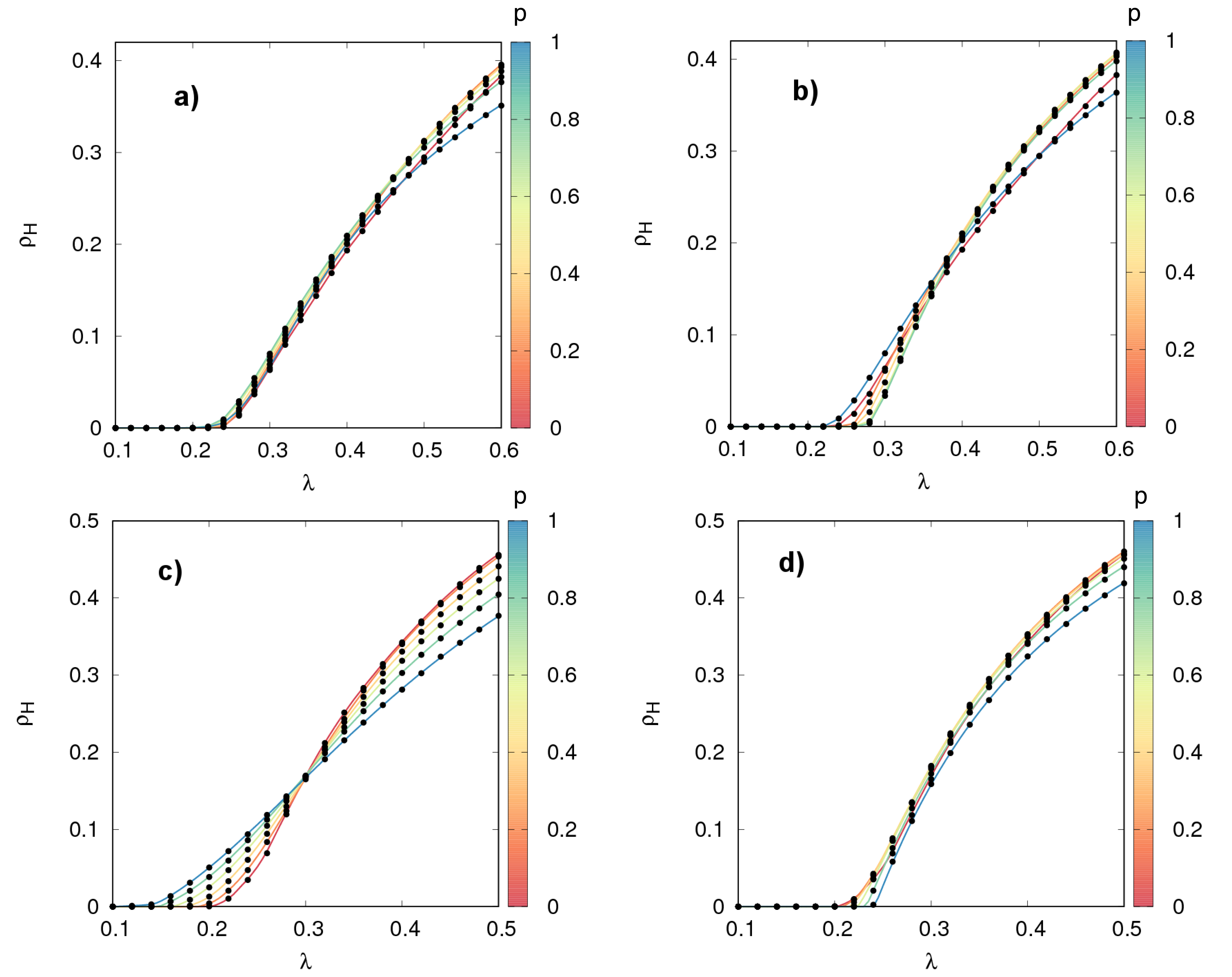}
\caption{Epidemic size $\rho_H$ as a function of the contagion rate between vectors and humans $\lm=\lh=\lambda$ and the human mobility $p$ (color code). Dots correspond to results of averaging 25 stochastic realizations whereas solid lines represent the theoretical predictions obtained by iterating the equations of the formalism. The recovery rate for both humans and vectors is set to $\mu^H=\mu^M=0.3$ Panels a)-b): The mobility network is the BA one described in the main text (see {\em Methods}). The values assumed for the restriction of the mobility of infected agents are (a) $\alpha=0$ and (b) $\alpha=1$. Panels c)-d): The  network used is the mobility network of Cali previously described in Sec.~\ref{sec:cali} Section. The values of infected agents mobility restriction parameter are (c) $\alpha=0$ and (d) $\alpha=1$.}
\label{FigMCMarkov}
\end{figure}

Let us first study the epidemic size --the fraction of the population who remains infected when the disease reaches a stationary state-- as a function of the contagion rates between humans and vectors $\lh,\lm$ and mobility $p$. To reduce the number of parameters and without loss of generality, let us define $\lm=\lh=\lambda$. We start analyzing the case in which human mobility is governed by an unweighted undirected Barab\'asi-Albert network (BA) of $N=50$ patches (see {\em Methods}), all of them homogeneously populated by $n_i = 1000$ agents. Concerning vectors' distribution, we consider that the ratio between vectors and humans populations inside a patch $i$, denoted in the following as $\gamma_i$, is randomly drawn from an uniform distribution within the range $\gamma_i \in [0.3,1.7]$. Figure \ref{FigMCMarkov} reveals the great agreement between theory and simulations for both the cases in which infected agents mobility is totally restrained, {\it i.e.} $\alpha=0$ (Fig. \ref{FigMCMarkov}a), and when there is no influence of the disease on agents' mobility, $\alpha=1$ (Fig. \ref{FigMCMarkov}b).

We now move to a real case, the city of Cali, whose metapopulation architecture is described in Sec.~\ref{sec:cali}. Following the same procedure, we can notice in Fig. \ref{FigMCMarkov}c-d that our formalism still captures very accurately the epidemic size, despite the higher complexity of Cali's mobility network.
Apart from the agreement between theory and simulations, interesting physical phenomena arise from accounting for the mobility constraints associated with contracting severe VBD. In particular, we notice that the qualitative behaviour of the epidemic threshold with human mobility can drastically change when modifying the restriction parameter. These phenomena will be analyzed in more detail in the next section where we compute the epidemic threshold as a function of ($p,\alpha$).

\begin{figure}[t!]
\centering
\includegraphics[width=0.90\columnwidth]{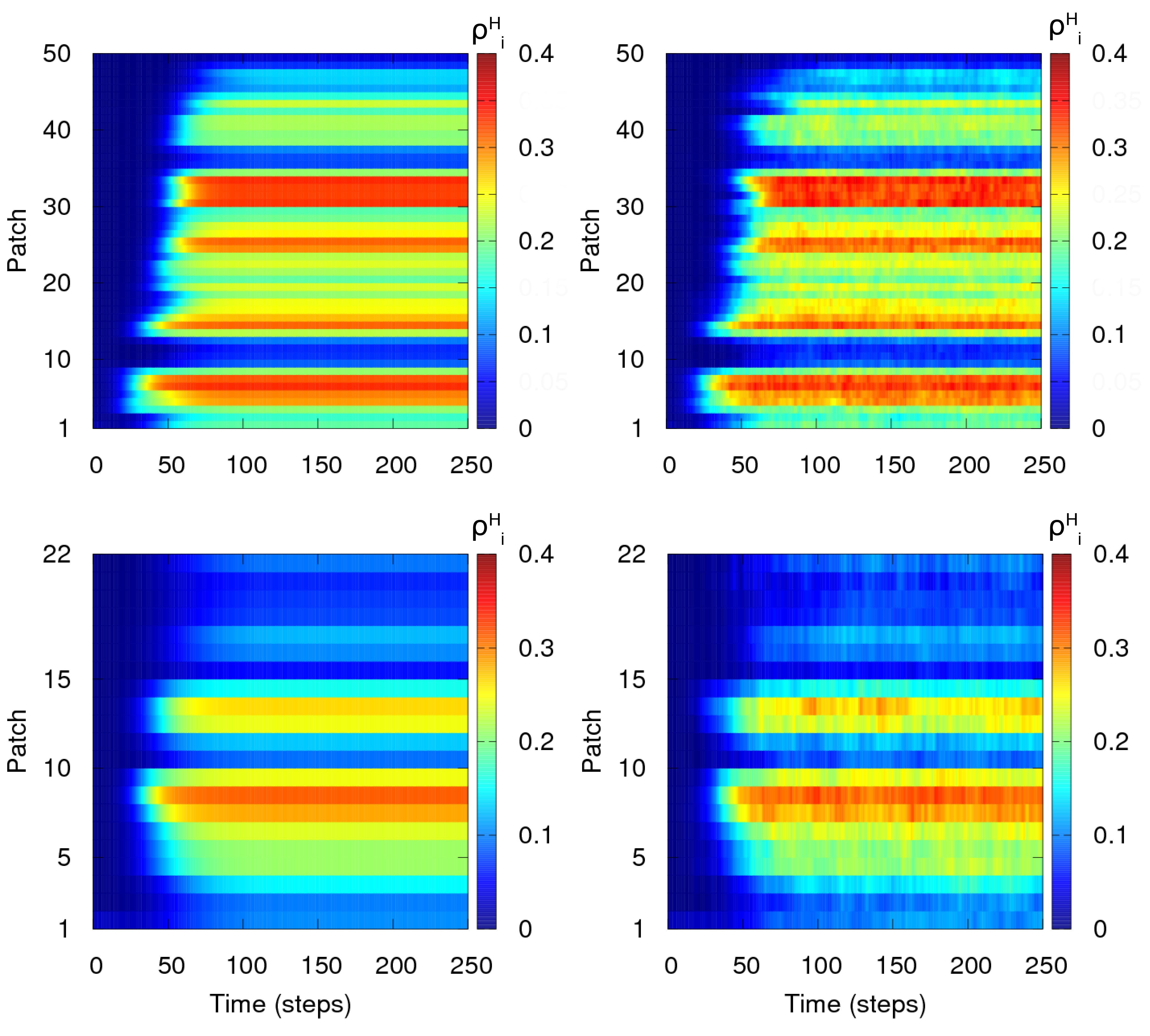}
\caption{Temporal evolution of the fraction of infected agents (color code) inside each patch. Parameters values are $(\lambda,\mu^H,\mu^M,p,\alpha) = (0.3,0.3,0.3,0.3,0)$. Top panels: the metapopulation which governs mobility processes is synthetic one described in the main text (see Methods). Bottom: The metapopulation used is the Cali mobility network detailed in the previous Section. The left panels correspond to the iteration of Eqs.(2-8) whereas the right ones contain the results of a single Monte Carlo realization.}
\label{Fig.temp}
\end{figure}

Finally, we also want our theory to reproduce the spatio-temporal unfolding patterns of VBD across both synthetic and real networks. To check this, we start by setting a seed localized in a single patch, and then we monitor the temporal evolution of the population affected by the disease inside each area. Figure \ref{Fig.temp} reveals that our formalism is able to capture the different propagation pathways of VBD in the BA and Cali metapopulations, despite the noise induced by the stochastic nature of the mechanistic simulations.

\section{Linearization of the Markovian Equations}
Here we derive an analytical estimation of the epidemic threshold for the Ross-Macdonald model with metapopulations. In the main manuscript we have defined this quantity as the minimum infection rate from vectors to humans which leads to an epidemic regime in the stationary state. We start considering the steady state of the Markovian equations. Assuming $\rho_i^H(t+1) = \rho_i^H(t) = \rho_i^H\ , \rho_i^M(t+1)=\rho_i^M(t) = \rho_i^M \ \forall i$, Eqs. (2,3) of the main text turn into:
\begin{eqnarray}
\mu^H\rho^{H}_{i} & = & (1-\rho^{H}_{i})I^{H}_i,\label{infected_h}\\
\mu^M\rho^{M}_{i} & = & (1-\rho^{M}_{i})I^{M}_i,\label{infected_m}
\end{eqnarray}
and the probabilities  $I^{H}$ and $I^{M}$ that a human or a vector associated with node $i$ becomes infected are respectively:
\begin{align}
 I^{H}_i &= (1-p)\left[1-\left(1-\lh\rho_{i}^{M}\frac{1}{n_{i}^{eff}(\rho_H,\alpha,p)})\right)^{\beta m_i}\right] \nonumber \\
  & \qquad + p\sum_{j=1}^{N}R_{ij}\left[1-\left(1-\lh\rho_{j}^{M}\frac{1}{n_{j}^{eff}(\rho_H,\alpha,p)})\right)^{\beta m_j}\right]\ .\label{IH}
 \end{align}
 \begin{eqnarray}
 I^{M}_i&=&1-\left(1-\lambda^{HM}\frac{(1-\alpha p)n_i\rho^{H}_{i}+\alpha p\sum_{j=1}^{N}R_{ji}\rho^{H}_j}{n_{i}^{eff}(\rho_H,\alpha,p)}\right)^{\beta}\label{IM}\ .
\end{eqnarray}

Close to the boundary between the epidemic and the disease-free phases, these steady values are small but not zero, which mathematically is reflected by considering $\rho_i^H = \epsilon_i^H\ll 1 \ ,\  \rho_i^M = \epsilon_i^M \ll 1 \ \forall i$. The latter allows us to linearize the equations by neglecting non-linear terms in $\epsilon$. Thus, Eqs.(\ref{IH},\ref{IM}) turn into:
\begin{eqnarray}
I^{H}_i &\simeq&(1-p)\lh\beta m_i\frac{1}{n_{i}^{eff}(0,\alpha,p)}\epsilon_{i}^{M}+p\sum_{j=1}^{N}R_{ij}\lh\beta m_j\frac{1}{n_{j}^{eff}(0,\alpha,p)}\epsilon_{j}^{M} . \label{epsilonH}\\
I^{M}_i&\simeq & \beta\lambda^{HM}\left[\frac{(1-\alpha p)n_i\epsilon^{H}_{i}}{n_{i}^{eff}(0,\alpha,p)}+\frac{\alpha p\sum_{j=1}^{N}R_{ji}n_j\epsilon^{H}_j}{n_{i}^{eff}(0,\alpha,p)}\right]\label{epsilonM}.
\end{eqnarray}

We now introduce the former expressions into Eqs. (\ref{infected_h},\ref{infected_m}) and keep only linear terms in $\epsilon$, yielding:
\begin{eqnarray}
\epsilon_i^H &=&\sum_{j=1}^N \frac{\lh\beta}{\mu_H}\underbrace{\left( pR_{ij}\frac{m_j}{\tilde{n}_j^{eff}}+(1-p)\delta_{ij}\frac{m_i}{\tilde{n}_i^{eff}} \right)}_{M_{ij}}\epsilon_j^{M} \label{epsilonH2}\ .\\
\epsilon_i^M &=&\sum_{j=1}^N \frac{\lm\beta}{\mu_M}\underbrace{\left(\alpha pR_{ji}\frac{n_j}{\tilde{n}_i^{eff}}+(1-\alpha p)\delta_{ij}\frac{n_i}{\tilde{n}_i^{eff}} \right)}_{\widetilde{M}_{ij}}\epsilon_j^{H}\ \label{epsilonM2} ,
\end{eqnarray}
where $\tilde{n}_i^{eff}$ has been defined as $\tilde{n}_i^{eff} = n_i^{eff}(0,\alpha,p)$. For the  sake of clarity, let us write the former system of equations in a more compact way:
\begin{equation}
\begin{pmatrix}
\vec{\epsilon}^{\;H}\\
\ \\
\vec{\epsilon}^{\;M}\\
\end{pmatrix}
= \begin{pmatrix}
0 & \frac{\beta\lh}{\mu^H}M \\\\ \frac{\beta\lm}{\mu^M}\widetilde{M} & 0
\end{pmatrix}
\begin{pmatrix}
\vec{\epsilon}^{\;H}\\
\ \\
\vec{\epsilon}^{\;M}\\
\end{pmatrix}\ .
\label{eq:matrix1}
\end{equation}

Equation \ref{eq:matrix1} makes evident the bipartite nature of the processes involved in the spread of VBD with matrices $M$ and $\widetilde{M}$ responsible for vector-human and human-vector infections, respectively. Thus, if we want to quantify indirect infections between humans mediated by vectors and vice versa we should iterate Eq.~\ref{eq:matrix1} obtaining:  
\begin{equation}
\begin{pmatrix}
\vec{\epsilon}^{\;H}\\
\ \\
\vec{\epsilon}^{\;M}\\
\end{pmatrix}
= \frac{\beta^2\lh\lm}{\mu^M\mu^H}\begin{pmatrix}
M\widetilde{M} & 0 \\\\ 0 & \widetilde{M}M 
\end{pmatrix}
\begin{pmatrix}
\vec{\epsilon}^{\;H}\\
\ \\
\vec{\epsilon}^{\;M}\\
\end{pmatrix}\ .
\label{eq:matrix2}
\end{equation}

From Eq.~\ref{eq:matrix2} it becomes clear that calculating the epidemic threshold involves solving an eigenvalue problem. In order to obtain a closed expression for the epidemic threshold and without loss of generality, we assume $\lh=\delta\lm$. Finally, as we are interested in the minimum vale of $\lm$ for which the former equation holds, the epidemic threshold can estimated as:
\begin{equation}
\lh_c = \sqrt{\frac{\mu_H\mu_M}{\delta\beta^2\Lambda_{max}({\bf M\widetilde{M}})}}\ ,
\label{Eq.eig}
\end{equation}
where $\Lambda_{max}$ is the spectral radius of matrix ${\bf M\widetilde{M}}$.

\section{Heuristic approximation of the epidemic threshold}
In the main text, we demonstrate that Eq. \ref{Eq.eig} accurately predicts the epidemic threshold. However, despite the accuracy of this estimation, its calculation involves computing the spectral radius of an arbitrarily large matrix which, for very extensive systems, can be a hard computational task. Here we derive a heuristic indicator that solves this problem and clarifies the intuition behind the different behaviors emerging as a result of human mobility. 

\begin{figure}[t!]
\centering
\includegraphics[width=0.95\columnwidth]{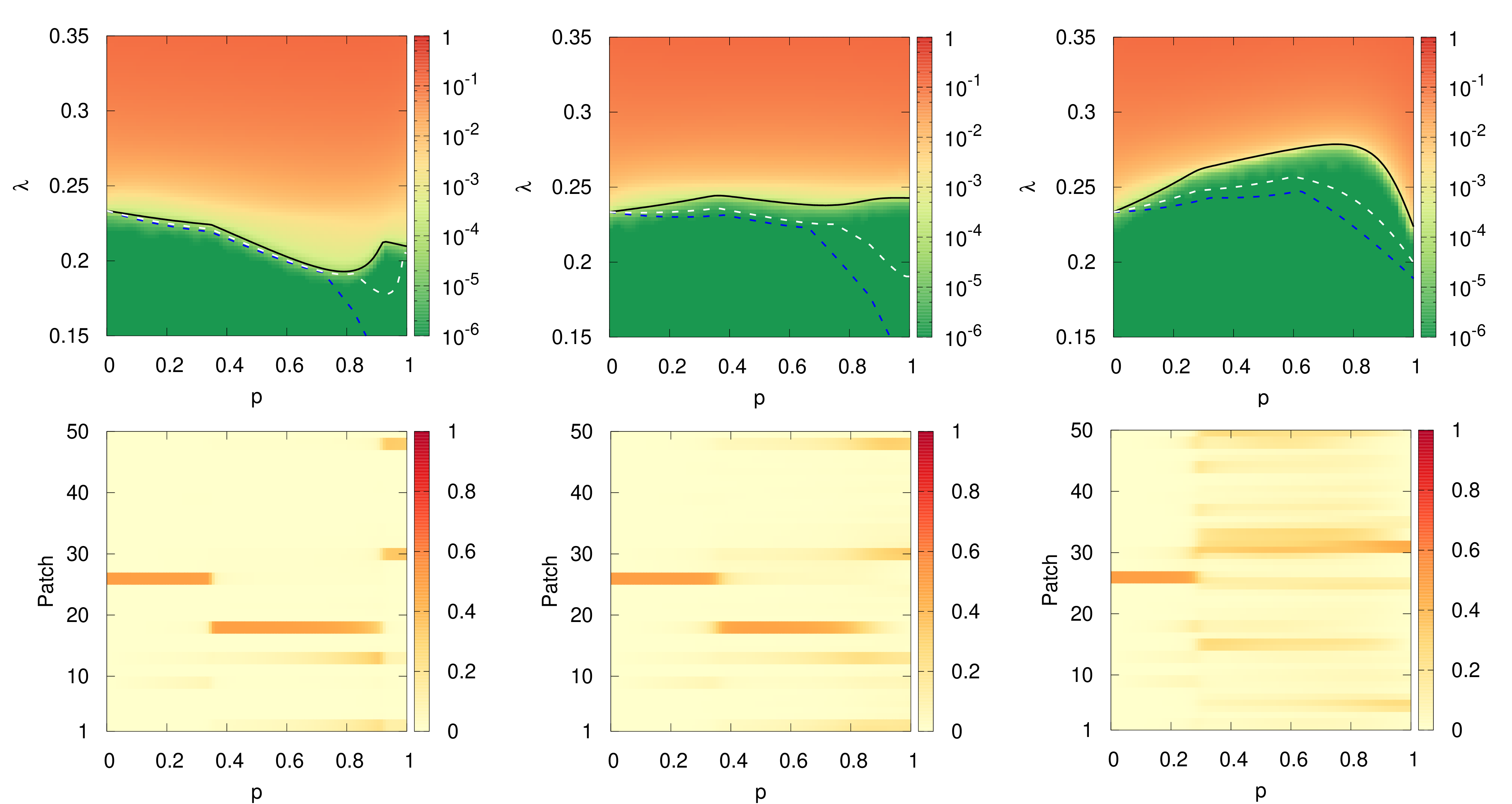}
\caption{Epidemic size (color code) as a function of the infectivity $\lh=\lm=\lambda$ and mobility $p$. From left to right values of $\alpha$ used are  $\alpha=0, 0.5, 1$. Solid black line corresponds to the theoretical prediction for the epidemic threshold given by Eq. \ref{Eq.eig}. Dashed lines represents the proposed indicators by taking into account 2-steps contagions (blue) and 4 steps ones (white). The remaining parameters have been set to $(\beta,\mu^H,\mu^M)=(1,0.3,0.3)$.}
\label{Fig.Eigs}
\end{figure}

Let us first discuss the physical meaning of Eqs. \ref{epsilonH} and \ref{epsilonM}. Eq. \ref{epsilonH} accounts for all the possible contagions processes affecting an agent with residence in $i$. Specifically, one agent associated with node $i$ can become infected in two possible ways: staying at node $i$ and getting bitten by a vector or moving to another sub-population and getting infected there. On the other hand, Eq. \ref{epsilonM} counts the infectious interactions of a vector from $i$ with agents that live there or belong to other patches and have decided to visit $i$. Thus, the number of indirect contagion processes $C^2_i$ received by a susceptible individual associated to a patch $i$ from infected agents can be quantified by composing these cross contagion rates across all the patches of the metapopulation, {\em i.e.}: 
\begin{equation}
C^2_i =\frac{1}{\mu_H\mu_M} \beta^2\delta(\lambda^{MH})^2\sum\limits_{j=1}^N\sum\limits_{k=1}^N M_{ik}\widetilde{M_{kj}}\ .
\end{equation}
Therefore, the larger this quantity for patch $i$ is, the higher the probability that an agent from $i$ contracts the disease. Following this line, we can thus approximate the epidemic threshold as the minimum value of $\lh$ needed to assure that an agent from the most exposed patch is infected at least once. This value, here defined as $r^2$, is given by:
\begin{equation}
r^2 = \sqrt{\frac{\mu_H\mu_M}{\delta\beta^2\max\limits_{i}(\sum\limits_{j=1}^N\sum\limits_{k=1}^N M_{ik}\widetilde{M_{kj}})}}.
\end{equation}
Fig. \ref{Fig.Eigs} shows a comparison between the epidemic threshold, calculated from the spectra of $M\widetilde{M}$ (black line), and $r^2$ (blue dashed line) as a function of human mobility $p$ for three different values of $\alpha$. From Fig. \ref{Fig.Eigs}, it can be observed that, regardless of  $\alpha$, $r^2$ approximates quite well the epidemic threshold for low mobilities, even the sharp variations due to the crossover between the most affected areas. However, $r^2$ becomes less accurate when the mobility of infected and susceptible individuals are quite different. 

To explain this discordance, we must take into account that, as infected individuals mobility is restrained, contagions between agents from different residences require more indirect contacts to happen. A very illustrative example is to consider two patches, say A and B, that are not mutually connected but both have connections to a third one, say C. If $\alpha=1$ an infected agent from A can pass the disease to a healthy individual from B in two steps:
\begin{enumerate}
\item{The infected agent moves to C and infects a vector there.}
\item{The infected vector bites a visitor from B, who becomes infected.}
\end{enumerate}
However, for $\alpha=0$, this cannot happen since infected individuals are not allowed to move. In fact, now the shortest contagion path contains 4 steps:
\begin{enumerate}
\item{The infected agent stays at A and a vector contracts the disease there.}
\item{A susceptible agent from C moves to A and becomes infected after being bitten.}
\item{This infected agent stays at C and transmits the disease to a vector there.}
\item{Finally, the infected vector at C passes the disease to the susceptible coming from B.}
\end{enumerate}
The latter example makes evident the necessity of considering contagion processes beyond two time steps to estimate the epidemic threshold. Following the same procedure as in the 2-step case, the number of indirect contagion processes that an agent from $i$ receives in 4 steps can be calculated by:
\begin{equation}
C^4_i = \frac{1}{\mu_H^2\mu_M^2}\beta^4\delta^2(\lambda^{MH})^4\sum\limits_{jklm}^N M_{ik}\widetilde{M_{kl}}M_{lm}\widetilde{M_{mj}}\ .
\end{equation}
Therefore, the minimum $\lh$ value to hold the epidemic, denoted as $r^4$, is given by:
\begin{equation}
r^4 = \sqrt[4]{\frac{\mu_H^2\mu_M^2}{\delta^2\beta^4\max\limits_{i}(\sum\limits_{jklm}^N M_{ik}\widetilde{M_{kl}}M_{lm}\widetilde{M_{mj}})}}
\end{equation}
Finally, Fig. \ref{Fig.Eigs} demonstrates that $r^4$ (white dashed line) provides a better estimation  for the threshold than  $r^2$, especially for hihg mobility values. As anticipated, this is mainly caused by the large number of contagious paths which are not captured by $C^2_i$ when the mobility of infected individuals is seriously constrained by the disease.
\begin{figure}[t!]
\centering
\includegraphics[width=0.88\columnwidth]{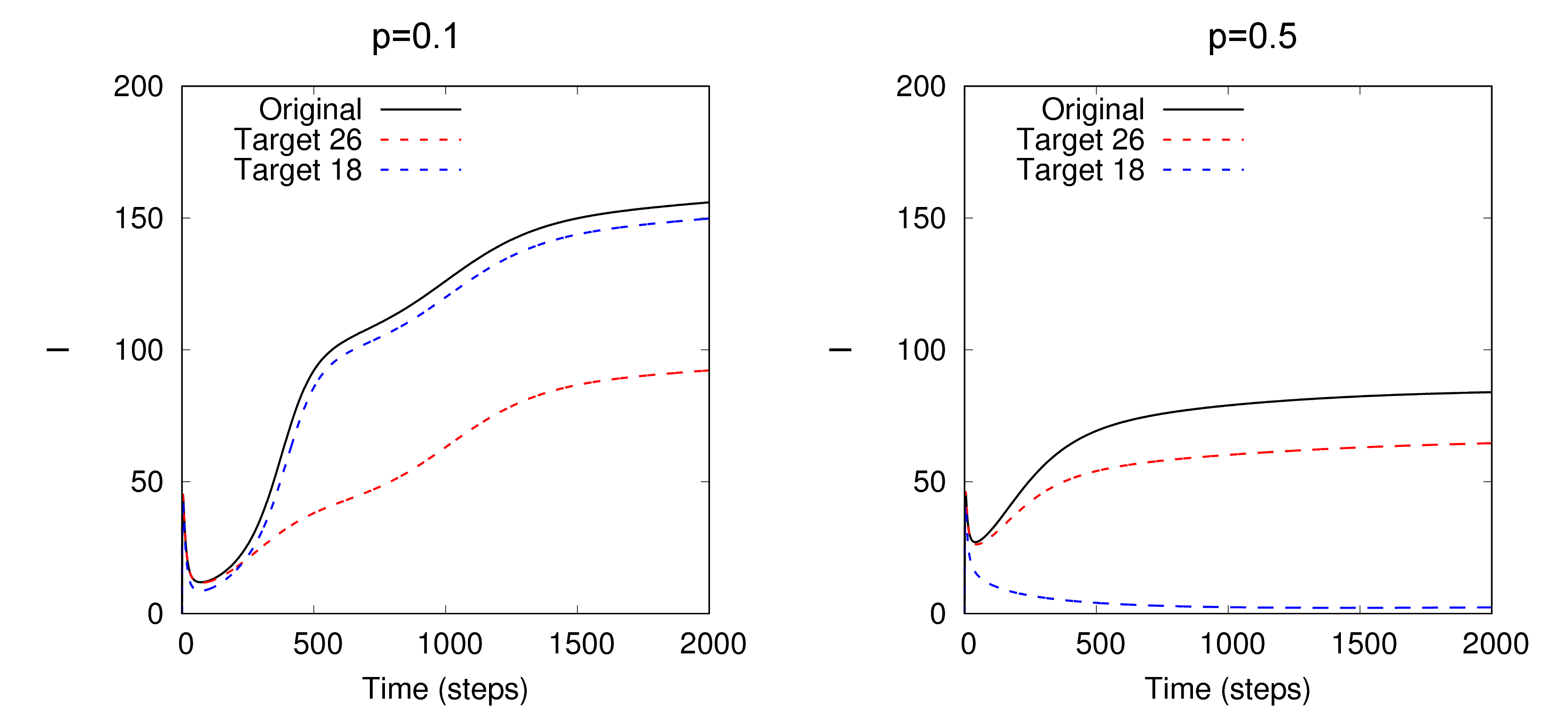}
\caption{Temporal evolution, according to the Markovian equations, of the number of infected agents by a VBD which spreads over  a BA network. The black line denotes the original curve where no policy has been implemented whereas dashed lines correspond to the case in which patches 26 (Red) and 18 (Blue) are immunized. The epidemic parameters have been set to $(\alpha,\lm,\lh,\mu^H,\mu^M)=(0.5,0.25,0.25,0.3,0.3)$. The values for human mobility are  $p=0.1$ ( $p<p_c$ Left panel) and $p=0.5$ ($p> p_c$ Right panel).}
\label{Fig:Immunization}
\end{figure}

\section{Consequences of the abrupt changes in the patches driving VBD}
The study of the critical properties of VBD in the main manuscript has revealed the existence of some mobility values, denoted as $p_c$, for which the components of the leading eigenvector change abruptly. Translated into epidemiological words, this striking phenomenon reflects the change in the most affected patches, which is of great relevance since targeted policies in specific areas can pass from useful to useless as human mobility varies. To prove it, we now study the effects of applying prevention measures in specific locations selected according to the largest components of the leading eigenvector of the critical matrix $\bf{M\tilde{M}}$. We consider the synthetic network used in the main text and set $\alpha=0.5$, for which the change in the leading patch happens at $p=p_c=0.38$ (see Fig.2 in main manuscript). Finally, we target two different sub-populations for the immunization:  patch 26 that is the leading patch for $p<p_c$, and patch 18 that sustains the outbreak for $p>p_c$.

Figure \ref{Fig:Immunization} confirms that the effectiveness of targeted policies against VBD is strongly influenced by aspects concerning human mobility. Thus, for example, immunizing agents from patch 18 leads to the extinction of  the disease for $p>p_c$, whereas it is almost ineffective for mobility values below this threshold. 
This result, along with the others presented in the main text, highlights the importance of promoting containment policies not only based on demography or entomological information but also taking into account the complex interplay between human movement, census data and vector abundance.

\begin{figure}[t!]
\centering
\includegraphics[width=0.85\columnwidth]{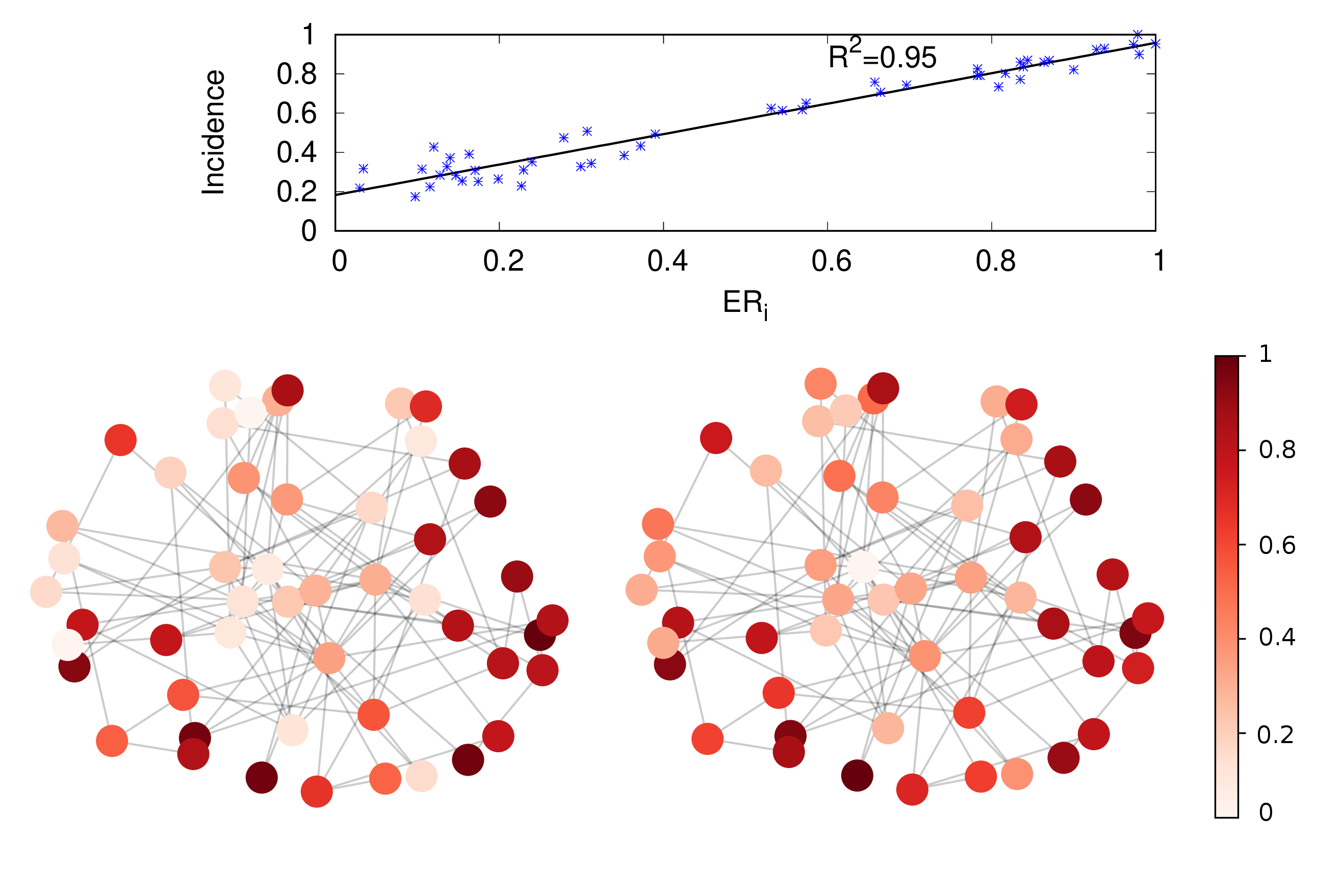}
\caption{{\it Upper panel:} correlation between the epidemic risk indicator $ER$ and the fraction of infected individuals in each patch given by the mechanistic simulations for a synthetic scale free network of $50$ patches with $n_i=10^3$ and $m_i \in [300,1700]$. {\it Lower panel:} spatial comparison. The color code represents: on the left, disease incidence obtained in the numerical simulations while on the right, the epidemic risk indicator computed according to Eq.~\ref{Eq.ER}. The parameters for the simulations have been set to $(p,\alpha,\beta,\lh,\lm,\mu^H,\mu^M) = (0.2,0.5,1,0.36,0.36,0.3,0.3)$ .}
\label{Fig.MCindicator}
\end{figure}
\section{Epidemic risk indicators: Synthetic and real metapopulations}
In the previous section, we proposed two heuristic indicators --$C_i^2$ and $C_i^4$-- able to capture, not only the dependence of the epidemic threshold on the mobility, but also changes in the patches driving the onset of epidemics. 
Here we go one step further and provide a prevalence indicator  able to reproduce the spatial distribution of the disease across sub-populations. Our aim is to classify the different geographical areas according to the risk associated to VBD. 
To do so, let us recall that $C^2_i$ encodes all the possible indirect infectious contacts that an agent from $i$ receives during an outbreak. Therefore, to obtain an estimate of the disease incidence in each patch, we must also account for the population of each geographical area. This way, we can define an epidemic risk indicator $ER_i$ for each patch $i$ as:
\begin{equation}
ER_i = n_i\sum\limits_{j=1}^N\sum\limits_{k=1}^N M_{ik}\widetilde{M_{kj}}
\label{Eq.ER}
\end{equation}

At this point, it is important to notice that Eq.~\ref{Eq.ER} does not depend on the epidemiological traits of the disease but only on human census, mobility and mosquitos abundance. In fact, as we assume all the epidemiological parameters to be constant across all the areas, dealing with one or another disease only leads to a rescaling of the epidemic risk of all the regions without producing any change in the ranking of affected populations.
\subsection*{Synthetic metapopulations}
We start testing the goodness of the predictions obtained via Eq.\ref{Eq.ER} in a controlled environment using a synthetic metapopulation.  We compare its value for each patch with the incidence of the disease from lenghty mechanistic simulations on the synthetic network used in the main text. We then fix the epidemiological parameters (namely: the cross-contagion rates $\lambda^{MH},\lambda^{HM}$, the vectors' biting rate $\beta$ and the recovery rates $\mu^H$ and $\mu^M$) as well as human mobility $p$ and $\alpha$ while human population inside each patch is constant ($n_i=10^3$) and vectors abundance is randomly assigned in the range $m_i \in [300,1700]$. Figure \ref{Fig.MCindicator} demonstrates the accuracy of our formalism in reproducing the unfolding of an outbreak in this controlled scenario, leading to a coefficient of determination $R^2 \simeq 0.95$.

\subsection*{Real metapopulations}
We  now check the applicability of our indicator to real scenarios. In particular, we focus on the spread of Dengue in the city of Cali in Colombia.  Dengue constitutes an important threat for Cali inhabitants so local authorities  monitor the situation and  publish  yearly  reports\cite{Dengue2015,Dengue2016}  with the spatial evolution of the disease across {\it comunas} and surveys about commuting inside the city are openly available\cite{surveysmed}.  

\begin{figure}[t!]
\centering
\includegraphics[width=1.0\columnwidth]{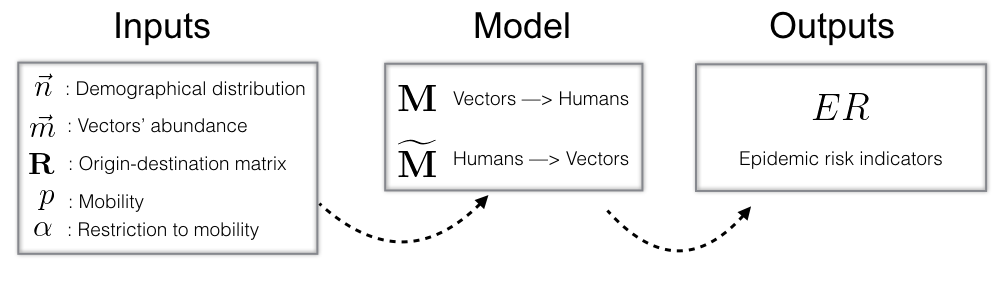}
\caption{Steps needed to compute the epidemic risk indicator for each geographical area $i$. First, it is necessary to construct a metapopulation network from the demographical distribution, the vectors abundance and the observed origin destination trips. Second, by fixing the value of the mobility $p$ as well as the restriction to the mobility of infected individuals $\alpha$, we can estimate the critical matrices ${\bf M}$ and ${\bf \widetilde{M}}$. Finally, we compute  the epidemic risk indicators $ER_I$  using Eq.~\ref{Eq.ER} }
\label{fig:steps}
\end{figure}

From the raw data only few steps are needed to compute the epidemic risk (see Fig~\ref{fig:steps}).  First, we should build the metapopulation systems. The inputs required are the demographical distribution $\vec{n}$, the vectors' abundance $\vec{m}$ and the recurrent mobility patterns encoded in the matrix ${\bf R}$.  Demographical distributions can be easily extracted from census published by local authorities -- in the case of Cali  we extracted them from reports issued by the Municipality\cite{mobility} -- while vectors' abundance can be inferred from entomological indexes associated to each area such as the recipient index or the pupae index. Finally, different methods have been proposed to estimate the origin-destination matrix  ${\bf R }$, which range from theoretical models\cite{Gonzalez2008} to the realization of surveys\cite{surveysmed} or the use of mobile phone call records\cite{Colizza2014}. 

With the structure and mobility of the metapopulation system fixed, we can compute the entries of matrices ${\bf M}$ and ${\bf \widetilde{M}}$, which  depend also on the parameters related to human mobility $\alpha,p$. To estimate $\alpha$ we recall that it represents the restriction in human mobility of infected individuals due to the effects of the disease. As symptoms associated with Dengue are usually severe, we can assume that all the symptomatic patients will not move, so $\alpha$ can be estimated as  the percentage of asymptomatic infections that are $\sim75\%$ of the infected population in case of Dengue disease.  In its turn, the mobility $p$ can be estimated as the fraction of individuals departing from their residence every day. Once defined the mobility, we can compute ${\bf M}$ and ${\bf \widetilde{M}}$ by using Eqs. \ref{epsilonH} and \ref{epsilonM} respectively. Finally, the epidemic risk is easily obtained by including these matrices in  Eq.~\ref{Eq.ER}.

The results of the comparison have been shown in Fig.~3 of the main text, where we compare the epidemic risk indicator with the spatial distribution of Dengue cases in Cali from 2015 to 2016. For this particular scenario, the mobility value $p$ is chosen to optimize the correlation between data and theory, yielding $p=0.36$. The good agreement between theoretical  indicator and the real cases ($R^2=0.81$) supports again the goodness of our approach, especially in presence of noisy and incomplete information, and its applicability to real world scenarios with the need of only few, and usually available, data. 

  \begin{figure}[t!]
 \centering
 \includegraphics[width=1.05\columnwidth]{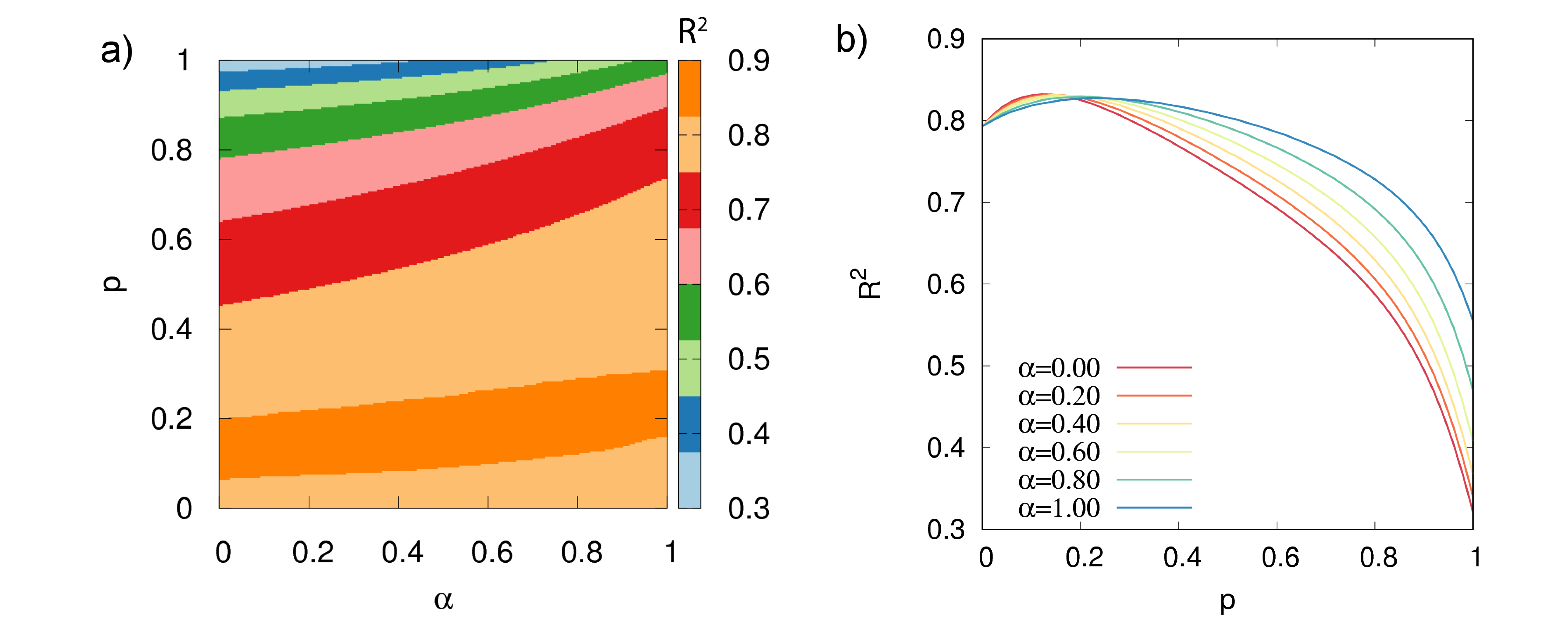}
 \caption{{\it(a)} Coefficient of determination $R^2$ (color code) as a function of the degree of mobility $p$ and the infected individuals mobility $\alpha$. {\it (b)} $R^2$ as a function of the mobility $p$ for several values of $\alpha$.}
 \label{Fig.Rob}
 \end{figure}
\section{Range of validity}
In the previous section, we have shown the accuracy of $ER_i$ while considering both real and synthetic metapopulations. However, as mentioned above, its value  depends on features related to human movements such as the degree of mobility $p$ or the disease induced constraint of the mobility $\alpha$. In this section we aim to test the robustness of $ER_i$ under variations of these parameters and, therefore, to determine its range of validity. 

To do so, we focus on the case of Cali and analyze the agreement between theory and the epidemiological data (quantified by $R^2$) as a function of both $p$ and $\alpha$. Fig.~\ref{Fig.Rob}a shows us  that $R^2$ remains over $0.80$ for mobility values below  $p=0.5$ and practically the entire range of $\alpha$. The agreement becomes more evident (see Fig.~\ref{Fig.Rob}b) if we consider larger values of $\alpha$ as in the case of Dengue ($\alpha \simeq 0.75$) where $R^2 \leq 0.75$ for values of human mobility up to $p=0.7$.   


\bibliographystyle{naturemag}

\end{document}